\documentclass[12pt]{article}
\pdfoutput=1
\usepackage{graphicx,epstopdf,amssymb,amsfonts,amsmath,amsthm,array,
mathrsfs,amscd}
\usepackage{todonotes} 
\newcommand{\pbs}[1]{\let\temp=\\#1\let\\=\temp}
\numberwithin{equation}{section}
\def\be{\begin{equation}}\def\ee{\end{equation}}
\def\cvp{\raise 2pt\hbox{,}}

 \def\d{{\rm d}}

\def\d{\partial}

\def\d{{\rm d}}

\def\a{\alpha}
\def\b{\beta}
\def\g{\gamma}
\def\G{\Gamma}
\def\dd{\delta}
\def\e{\epsilon}
\def\m{\mu}

\def\l{\lambda}
\def\r{\rho}
\def\s{\sigma}
\def\f{\phi}

\def\t{\theta}

\def\D{\Delta}

\def\vf{\varphi}

\def\wh{\widehat}
\def\wt{\widetilde}

\def\l{\lambda}

\def\del{\partial}

\def\cD{{\cal D}}
\def\cR{{\cal R}}

\def\ba{\begin{eqnarray}}
\def\ea{\end{eqnarray}}

\theoremstyle{plain}

\theoremstyle{definition}
\theoremstyle{remark}

\def\baselinestretch{1.30}

\begin{document}
%
%


{\pagestyle{empty}
\parskip 0in
\

\vfill
\begin{center}

{\Large \bf Small-time expansion of the Fokker-Planck kernel}
\vskip5.mm
{\Large \bf for space and time dependent diffusion}
\vskip5.mm
{\Large \bf and drift coefficients}

\vspace{0.4in}

{\large Adel B{\scshape ilal}} \\
\medskip
\it {
Laboratoire de Physique de l'\'Ecole Normale Sup\'erieure\\
\medskip
{\small ENS, Universit\'e PSL, 
CNRS,\\ 
Sorbonne Universit\'e,
Universit\'e Paris Diderot, Sorbonne Paris Cit\'e, Paris, France}
}

\smallskip

\end{center}
\vfill\noindent
We study the general solution of the  Fokker-Planck equation in $d$ dimensions with arbitrary space {\it and time} dependent diffusion matrix and drift term. We show how to construct the solution, for arbitrary initial distributions, as an asymptotic  expansion for small time. This generalizes the well-known asymptotic expansion of the heat-kernel for the Laplace operator on a general Riemannian manifold. We explicitly work out the general solution to leading and next-to-leading order in this small-time expansion, as well as to next-to-next-to-leading order for vanishing drift. We illustrate our results on a several examples.

\vfill
\medskip
%
\begin{flushleft}
\noindent
\end{flushleft}
\newpage\pagestyle{plain}

}

{\parskip -0.3mm
\newpage


\setcounter{page}{1}


%
%
\section{Introduction}

\subsection{Motivation}

When studying stochastic processes, a central equation is the Fokker-Planck equation, with its different versions also known as forward and backward Kolmogorov equations. An extensive discussion of this equation can be found in the book by Risken \cite{Risken}.

In the simple, one-dimensional case, the Fokker-Planck equation can be written as
\be\label{FPonedim}
\frac{\del}{\del t} \r(t,x)=\frac{\del^2}{\del x^2} [ D(t,x) \r(t,x)]-\frac{\del}{\del x} [ f(t,x) \r(t,x)] \ .
\ee
It corresponds to an Ito stochastic differential equation for a single random variable $X_t$ driven by a standard Wiener process $W_t$,
\be\label{SDEonedim} 
\d X_t=f(t,X_t) \d t + \s(t,X_t) \d W_t \ ,
\ee
if we identify $\r(t,x)$ as the probability density of the random variable $X_t$, and relate the diffusion coefficient $D$  to $\s$ through $D(t,x)=\s^2(t,x)/2$. The quantity $f$ is referred to as the drift or drift force. A standard reference is \cite{Gardiner}

From the physical point of view, the solution of the Fokker-Planck equation (in $d$ dimensions) describes a diffusion process, possibly  in the presence of a drift force.  For vanishing drift and constant diffusion coefficient $D$,  the solution corresponding to an initial distribution concentrated at a point $\vec{y}\ $ (i.e. $\r(0,\vec{x})=\dd^{(d)}(\vec{x}-\vec{y})$) is well-known:
\be\label{simplediffsol}
\r(t,\vec{x})=\frac{1}{(4\pi D\, t)^{d/2} }\exp\Big( - \frac{|\vec{x}-\vec{y}|^2}{4 D\, t} \Big) \ .
\ee
This shows immediately that the expectation value of the distance $|\vec{x}-\vec{y}|$ grows as $\sqrt{t}$, characteristic for a Brownian motion. 

In many simple situations the drift force may depend on  space and vary with time, while the diffusive behaviour is described by a single diffusion constant $D$ or, in a non-isotropic medium, by a constant diffusion matrix $D_{ij}$. More generally, in inhomogeneous media, the diffusion matrix may actually depend on the space-point. But the diffusive system described by the Fokker-Planck equation can be much more general, with the ``space coordinates" $x^i$ not  corresponding to any physical space, but to more general variables. Thus, in economics for example, the use of Fokker-Planck equations is rather widespread with the ``coordinates" corresponding to certain macro-economical variables, the diffusion matrix being implicitly determined from yet other data that, in particular, also depend on time (see e.g.~\cite{eco}). It is thus natural to study the Fokker-Planck equation with a very general space {\it and time} dependent diffusion matrix (and drift force).

Simple time dependent diffusion constants have much been considered to take into account certain anomalous diffusion behaviours \cite{JPBAG} in some (bio- or chemo-) physical systems (see e.g.~\cite{WU}). Also, solutions for certain simple time-dependences have been established,\footnote{
In particular, a time-dependence of the diffusion constant of the form $D(t)=\a\, t^{\a-1} \tilde D$ can be trivially removed by a change of variables $\tau=t^\a$, which then implies that the mean distance $|\vec{x}-\vec{y}|$ grows as $\sqrt{\tau}=t^{\a/2}$, a result referred to as anomalous diffusion.
}
but not much seems to be known about general time-dependent diffusion matrices.

\subsection{Outline of our approach}

The  general problem we consider in this note is to construct the solution of the general Fokker-Planck equation with an arbitrary initial condition, 
\be\label{FP1}
\frac{\del}{\del t} \r(t,x) = \cD_x \r(t,x) \quad , \quad \r(t_0,x)=\r_0(x) \ ,
\ee
where $x$ is short-hand for coordinates $x^1,\ldots x^d$ in a $d$-dimensional space and $\cD_x$ is a time-dependent second-order differential operator\footnote{Of course, $\cD$ is not to be confused with the diffusion constant.}
acting on $x$:
\be\label{Dxop}
\cD_x=G^{ij}(t,x)\del_i \del_j + B^i(t,x)\del_i +C(t,x) \ ,
\ee
where $\del_i=\frac{\del}{\del x^i}$. We use the convention that a repeated upper and lower index is summed (from 1 to $d$).  
For reasons to become clear, we have called the diffusion matrix $G^{ij}$ rather than $D_{ij}$,  the drift force is called $B^i$, and we have also added a ``scalar" term $C$.
As explicitly indicated, in addition to the space-dependence of the  coefficients $G^{ij}, B^i$ and $C$, we also allow them to vary with time. Obviously, $G^{ij}=G^{ji}$ and we will  require that it is positive and thus in particular non-degenerate. We also assume that the space described by the coordinates $x^i$ has no boundary,\footnote{
The fact of having no boundary has to be appreciated in the geometry determined by $G^{ij}$, as explained below, and not simply in terms of the coordinates.
 }
 so we do not need to provide any (spatial) boundary conditions.  Of course, with respect to time we have the initial condition $\r(t_0,x)=\r_0(x) $.
 In the following we will often not explicitly write the arguments $t$ and $x$ of the coefficients $G$, $B$ and $C$. The Fokker-Planck equation \eqref{FP1} with the differential operator written in the form \eqref{Dxop} is also referred to as the backward Kolmogorov equation.

To construct the solution of \eqref{FP1} for an arbitray initial distribution $\r(t_0,x)=\r_0(x)$ it is most convenient to find the corresponding ``Fokker-Planck kernel" or generalized heat kernel $K(t,x;t_0,y)$ satisfying
\be\label{FP2}
\frac{\del}{\del t} K(t,x;t_0,y) = \cD_x K(t,x;t_0,y) \ ,
\quad K(t,x;t_0,y)\sim e^{-V(t_0,y)}\, \dd^{(d)}(x-y) \ , \ \text{as}\ t\to t_0 \ ,
\ee
where $V(t,x)$ is some {\it a priori} arbitrary function so that
\be\label{measure}
\d\m_x(t)=e^{V(t,x)}\d^d x
\ee
defines an ``appropriate" measure on our space, at time $t$. Obviously then, the solution of \eqref{FP1}  is
\be\label{rhosol}
\r(t,x)=\int \d\m_y(t_0)\, K(t,x;t_0,y)\, \r_0(y) =\int \d^d y\, e^{V(t_0,y)}\, K(t,x;t_0,y)\, \r_0(y) \ .
\ee
Note, that one uses the integration measure at time $t_0$.
Let us insist that this provides the solution to the Fokker-Planck equation \eqref{FP1} with (arbitrary) initial condition $\r(t_0,x)=\r_0(x)$ for {\it any} choice of the real function $V$. We will see below that the most convenient choice will be in terms of the determinant of the diffusion matrix $G^{ij}$ at the initial time $t_0$:
\be\label{VG}
e^{-2V(t,x)}\equiv e^{-2V(x)}  =\det G^{ij}(t_0,x) \ ,
\ee
so that $V$ is time independent.\footnote{It might seem even more natural to choose the time dependent $e^{-2V(t,x)} =\det G^{ij}(t,x)$. While this would be just as good a choice and leads to the same set of equations \eqref{F0eq} and \eqref{Fneq}, their final rewriting as \eqref{F0eqbis} and \eqref{Fneqbis} would be  more complicated.}

In the literature one often finds the alternative form of the Fokker-Planck equation for $\r$ or $K$ with the second-order differential operator $\cD_x$ written as
\be\label{FPalter}
\frac{\del}{\del t} K = \cD_x K\quad , \quad \cD_x  K= \del_i \del_j \big(G^{ij} K\big) +\del_i\big( \wt B^i K\big) + \wt C  K\ .
\ee
This form is also called the forward Kolmogorov equation and it is equivalent to \eqref{FP1} or \eqref{FP2} with the coefficients simply related by
$B^i=\wt B^i+ 2 \del_j G^{ij}$ and $C=\wt C + \del_i \wt B^i + \del_i\del_j G^{ij}$. However, the most useful rewriting of the differential operator $\cD_x$ is to decompose it as\footnote{
The $\del_i$ in $\cD_x^{(1)}$ obviously is meant to act on everything to its right, including the function  on which $\cD_x^{(1)}$ is applied.
}
\be\label{Dxop2}
\cD_x=\cD_x^{(1)}+\cD_x^{(2)} \quad ,
\quad
\cD_x^{(1)}=e^{-V}\del_i(e^V G^{ij}\del_j )+C\quad ,
\quad
\cD_x^{(2)}=A^i\del_i \ , 
\ee
with
\be\label{Aidef}
A^i= \Big(B^i-\frac{\del  G^{ij}}{\del x^j}-G^{ij}\frac{\del  V}{\del x^j}\Big)\  .
\ee
Clearly, $\cD^{(1)}_x$ is a self-adjoint differential operator with respect to the measure  $\d\m=e^V \d^d x$. We see that $\cD_x$ can only be self-adjoint if $A_i=0$ so that $\cD_x^{(2)}$ vanishes, i.e.  if $B^i$ is chosen appropriately.\footnote{
To appreciate the physical meaning of this condition, note that for $G^{ij}=D\dd^{ij}$ (with constant $D$) this simply means that the drift force $B^i$ derives from a potential $\sim V$. 
} 

For self-adjoint $\cD_x$ quite powerful tools are available to study and constuct solutions of \eqref{FP1}. In particular,
for time-independent $G^{ij}$, $V$ and $C$, if $\l_n$ and $\vf_n(x)$ are the eigenvalues and real, orthonormalized eigenfunctions of the self-adjoint $\cD_x$, then the kernel $K$ is the usual heat kernel, given by (see e.g.~\cite{Gilkey}) 
\be\label{selfadjK}
K(t,x;t_0,y)=\sum_n e^{-(t-t_0)\l_n} \vf_n(x)\vf_n(y) \quad, \quad \text{for time-independent, self-adjoint $\cD_x$}\ ,
\ee 
which, of course, is time translation invariant. 
Note that for general time-dependent coefficients $G,\ A$ and $C$, the kernel $K$ obviously is not time-translation invariant, i.e.~$K(t,x;t_0,y)$ depends on $t$ and $t_0$ separately, and not just on the difference $t-t_0$. 

Much of the literature is concerned with the large-time asymptotics of the solutions of the Fokker-Planck equation and the question of whether and how a given initial solution $\r(t_0)$  tends to an equilibrium solution $\r_\infty$ as $t\to \infty$, see e.g.~\cite{MarVil}. This question is equivalent to establishing the large-time asymptotics of the kernel $K$. Clearly, for self-adjoint $\cD_x$ with time-independent coefficients, the large-time asymptotics is controlled by the smallest non-vanishing eigenvalue of $\cD_x$. But many interesting questions are less concerned with what happens at infinite time but, on the contrary, with the evolution on relatively small time scales. For this, one wants to know the small-time asymptotic expansion of the kernel $K$.

For the special case  $A^i=C=0$ and the choice \eqref{VG} for $V$, the differential operator $\cD_x=\cD_x^{(1)}$ is just the scalar Laplace operator on a Riemannian manifold with inverse metric tensor being $G^{ij}$. If moreover $G^{ij}$ is time independent, the corresponding heat kernel $K$
has a well-known asymptotic expansion for small time intervals $t-t_0$, given in terms of geometric expressions (geodesic distance, curvature tensors, etc), see e.g \cite{Vassil,BF}. This small-time expansion is based on a very physical intuition: for short time intervals, the ``particle" or the configuration described by the point $x$ cannot diffuse far from its original point $y$. Thus only the small-scale structure of the manifold on which the diffusion takes place can be important. But at small scales any Riemannian manifold looks  almost like flat space. Thus the heat kernel can be constructed as a ``perturbation" of the flat-space heat kernel which is  the well-known\footnote{
Following the mathematical literature, we consider times and distances as dimensionless quantities. Otherwise one would need to include a diffusion coefficient $D$ with units $\frac{{\rm m}^2}{{\rm s}}$ by replacing $t\to D t$, cf \eqref{simplediffsol}. Also, the form \eqref{flatK} implies that the coordinates are scaled such that for $x$ close to $y$ we have $G^{ij}\simeq \dd^{ij}$.
}
\be\label{flatK}
K_{\rm flat}(t,x;t_0,y)=\frac{1}{(4\pi (t-t_0))^{d/2}} \,\exp\Big(- \frac{(x-y)^2}{4(t-t_0)}\Big) \ ,
\ee
where $(x-y)^2=G_{ij} (x-y)^i(x-y)^j$ is the squared flat-space distance  between $x$ and $y$.

Our goal here is to similarly construct the asymptotic small-time expansion of the solution of \eqref{FP2} with general space {\it and} time dependent $\cD_x$ as given by  \eqref{Dxop}, \eqref{FPalter} or \eqref{Dxop2}. Even though this $\cD_x$ is not self-adjoint and less a Laplace operator on a Riemannian manifold, an underlying geometric picture will nevertheless be helpful as a guide for our construction. The reason is that the dominant short-time behaviour is governed by the two-derivative term $\sim G^{ij}\del_i\del_j$, while the other terms only are sub-dominant. Moreover, for the dominant short-time behaviour we may  replace $G^{ij}(t,x)$ by 
\be\label{metricdefGg}
G^{ij}(t_0,x)\equiv g^{ij}(x)\ ,
\ee 
which we can interpret as a standard inverse metric tensor on a Riemannian manifold. (As usual, we denote the metric tensor by $g_{ij}$ and its inverse by $g^{ij}$, so that $g_{ik}g^{kj}=\dd^j_i$.) Thus we expect that one can obtain an asymptotic expansion of $K$ as 
\be\label{Kexp}
K(t,x;t_0,y)=K_0(t,x;t_0,y) F(t,x;t_0,y) \ , \quad F(t,x;t_0,y)=\sum_{r\ge 0} (t-t_0)^r F_r(t_0,x,y) \ ,
\ee
with $F_0(t_0,y,y)=1$ and
\be\label{K0}
K_0(t,x;t_0,y)=\frac{1}{(4\pi (t-t_0))^{d/2}} \,\exp\Big( -\frac{\ell^2(x,y)}{4(t-t_0)}\Big) \ ,
\ee
where $\ell(x,y)$ is the geodesic distance\footnote{
We write $\ell^2(x,y)$ instead of $[\ell(x,y)]^2$.
} 
between $x$ and $y$, i.e. the length of the shortest path between $x$ and $y$ as measured with the metric $g_{ij}$ (which equals $(G^{-1})_{ij}$ at $t=t_0$). For the convenience of those readers who are not familiar with Riemannian differential geometry, we recall some basic notions\footnote{A pedestrian introduction to some of the geometric quantities involved and how they transform can also be found in chapter 4 of ref. \cite{Risken}.} 
in appendix A. 

In the following, to simplify our notations, we will assume that the origin of time is chosen such that
\be\label{tochoice}
t_0=0 \ ,
\ee
so that $t-t_0\to t$ and we suppress writing the dependence on the initial time $t_0$,  i.e.
\be\label{t0=0}
K(t,x;t_0,y)\to K(t,x,y) \quad , \quad F_r(t_0,x,y) \to F_r(x,y) \ , \quad \text{for $t_0=0$.}
\ee

Note that our ansatz \eqref{Kexp} and \eqref{K0} implies that the leading short-time behaviour of $K$ is given by $K_0$. Since $\ell^2(x,y)=g_{ij}(y)(x-y)^i(x-y)^j + {\cal O}((x-y)^3)$, one sees that $K_0$ vanishes exponentially unless $x-y$ is at most of order $\sqrt{t}$. Thus, the small $t$ expansion also is a small $(x-y)$ expansion. This implies that to leading order in $t$, we can simply replace $\ell^2(x,y)$ by $g_{ij}(y)(x-y)^i(x-y)^j = (G^{-1})_{ij}(0,y) (x-y)^i(x-y)^j$ and to this leading order $K(t,x,y)\simeq K_0(t,x,y)\simeq \frac{1}{(4\pi t)^{d/2}} \,\exp\Big( -\frac{(G^{-1})_{ij}(0,y)(x-y)^i(x-y)^j }{4t}\Big) \simeq K_{\rm flat}(t,x,y)$, which is a  well-known result, see e.g. in \cite{Risken}.

In this paper, in sect.~\ref{asymptoticexp}, we will establish an infinite hierarchy of differential equations for the small-time coefficients $F_r$  (sect~\ref{smalltexpan}) that, in principle, can be solved straighforwardly as an expansion in $(x-y)\simeq \sqrt{t}$ to any desired order (sect~\ref{Fnsolsec}). While the leading order (sect~\ref{leadingshorttime}) is given by $K_0$, in  sect~\ref{Fnsolsec1}   we will provide the explicit general form of the  next-to-leading order corrections  $\sim t \sim (x-y)^2$. This means, at this order,  we will determine $F_0(x,y)$ in an expansion around $y$ up to and including terms of order $(x-y)^2 \simeq \ell^2(x,y)$ and give $F_1(y,y)$. In sect~\ref{Fnsolsec2} we also work out the next-to-next-to-leading order corrections $\sim t^2\sim t(x-y)^2\sim(x-y)^4$ for the somewhat simpler case of vanishing drift force, and using normal coordinates, determining $F_0(x,y)$ to order $(x-y)^4\simeq \ell^4(x,y)$, $\ F_1(x,y)$ to order $(x-y)^2\simeq \ell^2(x,y)$ and $F_2(y,y)$. Then, in sect.~\ref{examples} we work out a few examples. Some of them are trivial in the sense that they can be solved exatly by some ``trick". This provides a rather non-trivial  check of our general results. We also provide a generic example to show how our formula  works in general. As already mentioned,  appendix \ref{Riemannian} gives some pragmatic introduction to  the notions of Riemannian geometry we use, appendix  \ref{app2}  provides some details on Riemannian normal coordinates, while in appendix \ref{app3}  we work out a few formulae related to the geodesic length needed in the main text.

\subsection{Some clarifying comments}

The reader only interested in our results may safely skip this sub-section.

At this point one might wonder whether it is not much easier to just iteratively solve the Fokker-Planck equation \eqref{FP1} for $\r(t,x)$. Indeed, one might just Taylor expand $\r(t,x)$ in $t$ around $t_0=0$ and use repeatedly \eqref{FP1} to obtain the higher derivatives, e.g. $\frac{\del^2}{\del t^2}\r=\frac{\del \cD_x}{\del t} \r + \cD_x^2 \r$, etc. This is equivalent to expressing the solution as the Dyson series using the time-ordering\footnote{
The time-ordering $T [ \cD_x(t_1) \ldots \cD_x(t_n)]$ is defined to yield the product of the differential operators $\cD_x(t_i)$ ordered with the operators having the larger time arguments to the left of those with the smaller arguments.
} 
$T$:
\be\label{Dysonrho}
\r(t,x)=T\Big[\sum_{n\ge 0} \frac{1}{n!} \Big( \int_{0}^t \d t' \cD_x(t') \Big)^n \Big]\r_0(x) \ .
\ee
Ultimately, if we are only interested in the solution $\r$ for one given initial condition $\r_0$, this is  equivalent\footnote{
To see how the two very different looking approaches are related, consider just the simplest case with $\cD_x=\dd^{ij} \del_i\del_j=\D$ the flat-space Laplace operator. Then $K$ is just the flat-space $K_{\rm flat}$ of \eqref{flatK} and to evaluate $\int\d^d y K(t,x,y) \r_0(y)=\int \d^d y \frac{1}{(4\pi t)^{d/2}} e^{-(x-y)^2/(4t)} \r_0(y)$ with a {\it smooth} $\r_0(y)$ one Taylor expands the latter around $x$ and performs the Gaussian integrations. This gives $\r_0(x) + t \D \r_0(x) + \frac{t^2}{2} (\D)^2 \r_0(x) + \ldots$. On the other hand, with the present time-independent $\cD_x$, the time-ordering in \eqref{Dysonrho} is irrelevant and it simply reads $\r(x)= e^{t\D} \r_0(x)$ which, upon expading in $t$, gives the same result.
} 
to  to our construction of the kernel $K(t,x,y)$ and evaluating $\int K(t,x,y) \r_0(y)\d\m_y$. However, if we  want to obtain the kernel $K$ itself, the corresponding Dyson series would read
\be\label {DysonK}
K(t,x,y)=T\Big[\sum_{n\ge 0} \frac{1}{n!} \Big( \int_{0}^t \d t' \cD_x(t') \Big)^n \Big]   \big( e^{-V(0,y)} \dd^{(d)}(x-y)\big)\ ,
\ee
giving $K$ as a sum of highly singular distributions. Formally, we could regard this as the Taylor expansion in $t$ of $K(t,x,y)$ around the singular  value $t=0$ with the higher coefficients more and more singular. These remarks should make clear why we did not persue this avenue but rather construct $K(t,x,y)$ as a perturbation around the well-defined (for $t\ne 0$) $K_0(t,x,y)$.

We end this introduction with some remarks on the geometric picture. First,
let us explain why it is very natural that the diffusion matrix $G^{ij}$ or $G^{ij}(t_0,x)=g^{ij}(x)$ plays the role of a an inverse metric. To make the argument simple, assume that $B^i=C=0$ and that $g^{ij}(x)=\g_i(x) \dd^{ij}$ is diagonal so that
\be\label{simplifiedeq}
\frac{\del}{\del t} \r=\Big(\sum_i \g_i(x) \frac{\del^2}{\del (x^i)^2}\Big) \r \ .
\ee
Then, for a ``particle" initially at $x$,  diffusion in the $i$-direction proceeds with an effective diffusion constant $\g_i(x)$ and, hence, in a given (infinitesimal) time interval $\dd t$, the particle will diffuse by a coordinate-distance $\dd x^i\sim \sqrt{\dd t\, \g_i(x)}$ in  the $i$-direction. Similarly, if the diffusion takes place in the $j$-direction, during the time interval $\dd t$ it will diffuse by a  coordinate-distance $\dd x^j\sim \sqrt{\dd t\, \g_j(x)}$. The reason for these different behaviours  can be reinterpreted by saying that the particle always diffuses by the same ``true" distance $\dd d\sim \sqrt{\dd t}$ in any direction and that the true distance in the $i$-direction at $x$ is $\frac{\dd x^i}{\sqrt{\g_i(x)}}$ and, similarly,  the true distance in the $j$-direction is $\frac{\dd x^j}{\sqrt{\g_j(x)}}$. Thus the true distance squared $(\dd d)^2$ between any two nearby points with coordinates $x^k$ and $x^k+\dd x^k$ is $(\dd d)^2=\sum_k \frac{(\dd x^k)^2}{\g_k(x)} $. Given our assumption about the diagonal form of $g^{ij}(x)$, its inverse $g_{ij}(x)=\frac{1}{\g_i(x)}\dd_{ij}$ is also diagonal and we can write for the true distance squared
\be\label{truedistsquared}
(\dd d)^2  =\sum_{i,j} g_{ij}(x) \dd x^i \dd x^j \ .
\ee
But this precisely is the definition\footnote{
Again, if we use dimensionless times and distances, the $\g_i$ are dimensionless, and so is the metric tensor. If, however, we measure times in seconds and distances in meters, then the $\g_i$ have units ${\rm m}^2/{\rm s}$. Thus the metric $g_{ij}$ has units ${\rm s}/{\rm m}^2$. This could be avoided by factoring explicitly some overall (inverse) diffusion constant ${D}^{-1}$.
} 
of a metric tensor $g_{ij}$: it tells us what is the true distance-squared between the point with coordinates $x^i$ and the infinitesimally close point with coordinates $x^i+\d x^i$.

In many applications, this geometric point of view will be useful, if not necessary. The coordinates $x^i$ are certain parameters on which the distribution $\r$ depends and one might want to reformulate the problem in terms of new parameters $x^{'i}$ that are defined as appropriate functions of the old ones: $x^{'i}=f^i(x^j)$. This is a general coordinate transformation and it is then necessary to know how the coefficients of the differential operator $\cD_x$ in \eqref{FP1} and \eqref{Dxop2} transform.
We already mentioned that $g^{ij}$ is an inverse metric and $G^{ij}$ indeed transforms as a tensor,  while (with $e^{-2V}=\det_{ij} G^{ij}$), the $A^i$ transform as a vector:\footnote{
Note that it is $A^i$ that transforms as a vector and not $B^i$. This remains equally valid with our time-independent choice $e^{-2V(t,x)}=\det_{ij} G^{ij}(t_0,x)$.
}
\be\label{cordinatetransf}
G^{'kl}(t,x')=G^{ij}(t,x)\, \frac{\del x^{'k}}{\del x^ii} \frac{\del x^{'l}}{\del x^j} \ , \quad
A^{'k}(t,x')=A^i(t,x)\frac{\del x^{'k}}{\del x^i} \ .
\ee
\vskip2.mm

We have already argued why the diffusion matrix $G^{ij}(t,x)$ at fixed $t$ should be interpreted as an inverse metric on a space with coordinates $x^i$, turning the  diffusion problem on flat ${\bf R}^d$ into a geometric problem on  an a priori curved Riemannian manifold. This latter could even have a non-trivial topology, as can be   seen on the following simple example where the diffusion actually takes place on the sphere. Suppose we have just 2 coordinates $x^1$ and $x^2$ in the plane ${\bf R}^2$, and the diffusion matrix is time-independent and equals $G^{ij}(x)=g^{ij}(x)=[1+(x^1)^2+(x^2)^2]^2\ \dd^{ij}$. Then $g_{ij}=\frac{1}{[1+(x^1)^2+(x^2)^2]^2}\ \dd^{ij}$ and the infinitesimal distance $\d s$ between two points of coordinates $x^i$ and coordinates $x^i+\d x^i$ is given by $\d s^2=g_{ij}\d x^i \d x^j=\frac{(\d x^1)^2 + (\d x^2)^2}{[1+(x^1)^2+(x^2)^2]^2}$. If instead of $x^1$ and $x^2$ we use polar coordinates $r$ and $\f$ on the plane we get $\d s^2=\frac{\d r^2 + r^2 \d\f^2}{(1+r^2)^2}$.
It is easy to show that this exactly corresponds to the standard two-dimensional sphere described by the stereographic projection\footnote{
The stereographic projection is obtained by ``posing" the south-pole of the two-sphere on the origin of the plane and imagine a straight line through the north pole and any given point $P$ of the sphere. The coordinates of the point where this line intersects the plane are  the stereographic coordinates of the point $P$. This works for all $P$ except the northpole which is projected ``to infinity".
}
on the plane, with $r=0$ corresponding to the south pole and $r\to\infty$ to the north pole. It is the singularity of the  metric at $r\to\infty$ ($g_{ij}$ vanishes and $g^{ij}$ diverges) which makes it possible to describe the compact sphere by the non-compact plane. Physically, this singularity  corresponds to a diverging diffusion matrix.

Finally, let us come back to our assumption that the space has no boundary. This has to be appreciated in the geometry as determined by the metric $g_{ij}$. In the previous example, the ``boundary at $r=\infty$" just corresponds to an ordianry point on the sphere. As another example, consider $g_{11}=1$,  $g_{22}=(x^1)^2$, $g_{12}=g_{21}=0$ with $0\le x^1$ and $0\le x^2 <2\pi$ and periodic identification of $x^2$  with $x^2+2\pi$. Then $x^1=r$ and $x^2=\phi$ are just standard polar coordinates on the plane and there is no boundary contrary to what the condition $x^1=r\ge 0$ might have suggested.

\newpage

\section{The asymptotic expansion\label{asymptoticexp}}

We will now construct the asymptotic small-time expansion of the Fokker-Planck kernel $K(t,x,y)\equiv K(t,x;0,y)$ corresponding to the general equation \eqref{FP2} with $\cD_x$ given by \eqref{Dxop} (or \eqref{FPalter} or \eqref{Dxop2}) and an appropriately chosen  $V$ given below. Recall that $G_{ij}(t,x)$ is the inverse matrix of $G^{ij}(t,x)$ and $g_{ij}(x)=G_{ij}(t_0,x)\equiv G_{ij}(0,x)$. Moreover, $\ell(x,y)$ denotes the geodesic distance  between $x$ and $y$ as measured with the metric $g_{ij}$.

\subsection{The leading short-time behaviour\label{leadingshorttime}}

To begin with, we will justify that \eqref{Kexp} and \eqref{K0} provide the correct form of the asymptotic expansion we are looking for. In particular, we must show that $K_0$ provides the correct leading small-$t$ solution to \eqref{FP2}. Clearly, the leading term generated by taking $\frac{\del}{\del t}$ is
\be\label{ddtKleading}
\frac{\del}{\del t} K(t,x,y) = K_0(t,x,y) F_0(x,y) \Big[ \frac{\ell^2(x,y)}{4t^2}  +{\cal O}\big(\frac{1}{t}\big)\Big] \ ,
\ee
while the leading term generated by taking $\cD_x$ is
\be\label{DxKleading}
\cD_x K(t,x,y) = K_0(t,x,y) F_0(x,y) \Big[G^{ij}(t,x) \frac{\del_i \ell^2(x,y)}{4t}\frac{\del_j \ell^2(x,y)}{4t}+{\cal O}\big(\frac{1}{t}\big)\Big] \ ,
\ee
Now, the geodesic length satisfies
\be\label{georel}
g^{ij}(x) \del_i \ell^2(x,y) \del_j\ell^2(x,y) = 4 \ell^2(x,y) \ ,
\ee
where, as always, $\del_i = \frac{\del}{\del x^i}$. The simplest way to prove this relation is to note that it is coordinate-independent and thus it is enough to prove it in any convenient coordinate system. A particularly convenient choice are Riemann normal coordinates centered in $y$. The normal coordianates $\wt x^i$ then are defined by the geodesics through the point $y$, see App.~\ref{Riemannian}. It follows that $\ell^2(\wt x,\wt y)= \sum_i \wt x^i \wt x^i$ and $\wt g^{ij}=\dd^{ij}+ {\cal O}((\wt x^k)^2)$. All the ${\cal O}((\wt x^k)^2)$-terms have antisymmetry properties such that they vanish when multiplied with $\wt x^i$ or $\wt x^j$ and summed over $i$ or $j$ so that $\sum_j \wt g^{ij} \wt x^j=\wt x^i$. The relation \eqref{georel} then trivially follows in these coordinates and thus is always true.
Finally, $G^{ij}(t,x)=g^{ij}(x) + {\cal O}(t)$ and, using \eqref{georel}, eq.~\eqref{DxKleading}
becomes
\be\label{DxKleading2}
\cD_x K(t,x,y) = K_0(t,x,y) F_0(x,y) \Big[\frac{\ell^2(x,y)}{4t^2}+{\cal O}\big(\frac{1}{t}\big)\Big] \ ,
\ee
thus matching the leading term in \eqref{ddtKleading}.

In the limit $t\to 0$, the kernel $K(t,x,y)$ should become a $d$-dimensional Dirac distribution $\dd^{(d)}(x-y)$ times some function $e^{-V(x)}$, cf~\eqref{FP2}, which allows us to identify the appropriate measure for integrating in \eqref{rhosol} as $\d\m=e^{V(x)}\, \d^d x$. The basic formula we use is
$\frac{e^{-z^2/\e}}{\sqrt{\pi\e}} \sim \dd(z)$ as $\e\to 0$, and its $d$-dimensional generalization
\be\label{dirac2}
\frac{e^{-M_{ij}z^i z^j/\e}}{(\pi\e)^{d/2}} \sim \frac{1}{\sqrt{\det M}} \prod_{i=1}^d \dd(z^i)\equiv  \frac{1}{\sqrt{\det M}} \dd^{(d)}(z)
 \quad , \quad \text{as}\ \e\to 0\ ,
\ee
Now, as $t\to 0$, the exponential in $K_0$ will be arbitrarily small unless $x\to y$. In this limit, $\ell^2(x,y) \sim g_{ij}(y) (x-y)^i(x-y)^j$ and,  using the previous formula, as well as $F_0(y,y)=1$, immediately shows that
\be\label{K0limit}
K(t,x,y)\sim K_0(t,x,y)\sim \frac{1}{\sqrt{g(y)}}\, \dd^{(d)}(x-y) 
\quad , \quad \text{as}\ t\to 0 \ ,
\ee
where we have used the standard notation that $g$ denotes the determinant of the metric:
\be\label{detg}
g=\det g_{ij} \ .
\ee
Thus we identify
\be\label{Vrelg}
e^{V(0,x)}=\sqrt{g(x)} \ ,
\ee 
consistent with \eqref{VG}.
It is  satisfying to find that at the initial time the integration measure $\d\m_x(t)=e^{V(t,x)}\d^d x$, as defined in \eqref{measure}, turns out to be the standard volume element on the Riemannian manifold with metric $g_{ij}$:
\be\label{measure2}
\d\m_x(0)=\sqrt{g(x)}\,\d^d x \ .
\ee

Let us insist that $\r(t,x)$ as given by \eqref{rhosol} satisfies the Fokker-Planck equation $\frac{\del \r}{\del t} =\cD_x \r$ with initial condition $\r(t,x)=\r_0(x)$ for any choice of $G^{ij}$, $B^i$ and $C$, and {\it any} real function $V$ and thus the choice \eqref{Vrelg} does not restrict the generality of this solution in any way.

\subsection{The equations of the small-$t$ expansion\label{smalltexpan}}

We now insert the ansatz \eqref{Kexp} and \eqref{K0} (with $t_0=0$) into the Fokker-Planck equation \eqref{FP2} for the kernel $K$. This results in the following differential equation for $F(t,x,y)$:
\be\label{FPF}
\frac{\del}{\del t} F
=\Big( \frac{G^{ij} \del_i\ell^2 \del_j\ell^2-4\ell^2}{16 t^2} 
-\frac{G^{ij}\del_i\del_j\ell^2 +B^i\del_i\ell^2 -2 d}{4t}\Big) F
-\frac{G^{ij}\del_i\ell^2}{2 t}\, \del_j F + \cD_x F \ .
\ee
Of course, this looks much more complicated than the initial Fokker-Planck equation for $K$, but contrary to $K$, the function $F$ is required to be regular as $t\to 0$. If we want to rewrite this in a way which is manifestly invariant under transformations of the cordinates $x^i$ we must express $B^i$ in terms of $A^i$, cf~ eq.~\eqref{Dxop2}, since the $A^i$, not the $B^i$, transform in a well-defined way. The difference of the two terms should turn $G^{ij}\del_i\del_j\ell^2$ into $G^{ij}\nabla_i\del_j\ell^2$ where $\nabla_i$ is the covariant derivative for the metric $G$. For the time being, we keep the somewhat simpler looking equation \eqref{FPF}, but we will come back to this point below. 
We expand the coefficients appearing in $\cD_x$ as
\ba\label{coeffexp}
G^{ij}(t,x)&=& g^{ij}(x)+\sum_{r\ge 1} t^r g_{(r)}^{ij}(x)\equiv g^{ij}(x)+t\wh G^{ij}(t,x) \ ,
\nonumber\\
B^{i}(t,x)&=& b^i(x)+\sum_{r\ge 1} t^r b_{(r)}^{i}(x)\equiv b^{i}(x)+t\wh B^{i}(t,x) \ ,
\nonumber\\
C(t,x)&=& c(x)+\sum_{r\ge 1} t^r c_{(r)}(x)\equiv c(x)+t\wh C^{ij}(t,x) \ ,
\ea
and also
\be\label{Aexp}
A^{i}(t,x)= a^i(x)+\sum_{r\ge 1} t^r a_{(r)}^{i}(x) \ .
\ee
Note that, if we had not set $t_0=0$, this would be an expansion in powers of $t-t_0$ and the coefficients $g_{(r)}^{ij}(x), \ b_{(r)}^{i}(x),\ c_{(r)}(x)$ and $a_{(r)}^{i}(x)$ of course would also depend on the initial time $t_0$.
Note also that we do {\it not} require the $g^{ij}_{(r)}$ for $r\ge 1$ to be non-degenerate or non-negative.
Inserting the expansions \eqref{coeffexp} into \eqref{FPF} and, using again \eqref{georel}, we see once more that the leading $\frac{1}{t^2}$-term cancels:
\be\label{leadingvanishes}
\frac{G^{ij} \del_i\ell^2 \del_j\ell^2-4\ell^2}{16 t^2}=\frac{g^{ij} \del_i\ell^2 \del_j\ell^2-4\ell^2}{16 t^2}+ \frac{\wh G^{ij} \del_i\ell^2 \del_j\ell^2}{16 t}=0+\frac{\wh G^{ij} \del_i\ell^2 \del_j\ell^2}{16 t} \ .
\ee
Thus \eqref{FPF} can be rewritten as
\be\label{FPF2}
\frac{\del}{\del t} F
=\Big(
\frac{\frac{1}{4}\wh G^{ij} \del_i\ell^2 \del_j\ell^2 - G^{ij}\del_i\del_j\ell^2 +2d -B^i\del_i\ell^2}{4t} \Big) F
-\frac{G^{ij}\del_i\ell^2}{2 t}\, \del_j F + \cD_x F \ .
\ee
Equating the coefficients of the powers of $t$ results in a system of differential equations for the $F_r$. The terms $\sim\frac{1}{t}$ result in an equation for $F_0$ only:
\be\label{F0eq}
2 g^{ij}\del_i\ell^2\, \del_j F_0 = \Big( \frac{1}{4}g_{(1)}^{ij} \del_i\ell^2 \del_j\ell^2 - g^{ij}\del_i\del_j\ell^2 +2d -b^i\del_i\ell^2 \Big) F_0  \ ,
\ee
while the terms $\sim t^n,\ n\ge 0$ in \eqref{FPF2} provide equations for $F_{n+1}$ with inhomogeneous terms involving the $F_{r}$ with $0\le r\le n$:
\ba\label{Fneq}
&&\hskip-3.cm 
\Big( 4n+4 -\frac{1}{4}g_{(1)}^{ij} \del_i\ell^2 \del_j\ell^2 + g^{ij}\del_i\del_j\ell^2 -2d 
+b^i\del_i\ell^2 +2 g^{ij}\del_i\ell^2\, \del_j \Big)  F_{n+1}
\nonumber\\
&=&
\sum_{r=0}^{n} \Big(  \frac{1}{4}g_{(r+2)}^{ij} \del_i\ell^2 \del_j\ell^2 
-\ g_{(r+1)}^{ij}\del_i\del_j\ell^2 - b_{(r+1)}^i\del_i\ell^2
-2g_{(r+1)}^{ij} \del_i\ell^2 \del_j 
\nonumber\\
&&\hskip0.8cm + 4\big( g_{(r)}^{ij} \del_i \del_j +b_{(r)}^i \del_i +c_{(r)}\big) \Big) F_{n-r}  
\quad , \hskip3.5cm n\ge 0 \ .
\ea
In particular, for $n=0$, this is a differential equation for $F_1$ with the inhomogeneous term on the right-hand side involving only $F_0$. Recall that the normalisation  is fixed by the ``initial condition" $F_0(y,y)=1$.

Let us now replace the $b^i_{(r)}$ by the $a^i_{(r)}$. Expanding \eqref{Aidef} in powers of $t$ and using the fact that $V=\log \sqrt{g}$ (not $\log\sqrt{G}$ !) is time-independent, yields the relations
\be\label{aibi}
b^i_{(r)}=a^i_{(r)} +\frac{1}{\sqrt{g}}\del_j \big(g^{ij}_{(r)}\sqrt{g}\big) \ ,
\ee
which for $r=0$ can also be written in terms of the Christoffel symbol as 
\be\label{abrelGamma}
b^i=a^i -g^{kl}\G^i_{kl}\ . 
\ee
Introducing 
\be\label{Laplaciansr}
\D=\frac{1}{\sqrt{g}} \del_i (\sqrt{g}g^{ij}\del_j) = g^{ij}\del_i\del_j -g^{ij}\G_{ij}^k \del_k\quad , \quad
\D_{(r)}=\frac{1}{\sqrt{g}} \del_i (\sqrt{g}g_{(r)}^{ij}\del_j )\ ,
\ee
where $\D$ is just the scalar Laplacian on the manifold with metric $g_{ij}$ and the $\D_{(r)}$ are some other second-order scalar differential operators,\footnote{
For any scalar function $f$ we have $\D_r f= \del_i (g^{ij}_{(r)} \del_j f) +\frac{\del_i\sqrt{g}}{\sqrt{g}} g^{ij}_{(r)}\del_j f= \del_l (g^{lj}_{(r)} \del_j f) +\G^l_{li} g^{ij}_{(r)}\del_j f=\nabla_l (g^{lj}_{(r)} \del_j f)$ which manifestly is a scalar quantity.
}
we can then rewrite \eqref{F0eq} and \eqref{Fneq} as
\be\label{F0eqbis}
2 g^{ij}\del_i\ell^2\, \del_j F_0 = \Big( \frac{1}{4}g_{(1)}^{ij} \del_i\ell^2 \del_j\ell^2 -(\D\ell^2) +2d -a^i\del_i\ell^2 \Big) F_0  \ ,
\ee
with $F_0(y,y)=1$, 
and, for $n\ge 0$
\ba\label{Fneqbis}
&&\hskip-0.5cm 
\Big( 4n+4 -\frac{1}{4}g_{(1)}^{ij} \del_i\ell^2 \del_j\ell^2 + (\D\ell^2) -2d 
+a^i\del_i\ell^2 +2 g^{ij}\del_i\ell^2\, \del_j \Big)  F_{n+1}
\nonumber\\
&&\hskip-0.5cm=
\sum_{r=0}^{n} \Big(  \frac{1}{4}g_{(r+2)}^{ij} \del_i\ell^2 \del_j\ell^2 
-(\D_{(r+1)}\ell^2)- a_{(r+1)}^i\del_i\ell^2
-2g_{(r+1)}^{ij} \del_i\ell^2 \del_j 
+ 4\,\big( \D_{(r)}+a_{(r)}^i \del_i +c_{(r)}\big) \Big) F_{n-r}   \ .
\nonumber\\
\ea
It is maybe useful to recall that the $F_r\equiv F_r(x,y)\equiv F_r(t_0,x,y)$ depend on the initial time $t_0$ since all quantities appearing in these equations are determined through the $g^{ij}(x)$, $a^i(x)$, $c(x)$, as well as the $g_{(r)}^{ij}(x)$, $a_{(r)}^i(x)$, $c_{(r)}(x)$  which in turn are determined by the expansions of $G^{ij}(t,x)$, $A^i(t,x)$, $C(t,x)$ in powers of $t-t_0$ around the initial time $t_0$. Obviously, the equations for the $F_r$ are valid for arbitrary $t_0$, not just $t_0=0$.
Obviously also, the Laplace operators $\D,\ \D_{(r)}$ and partial derivatives $\del_i$ all act on $x$, not $y$. 
Note that the equations \eqref{F0eqbis} and \eqref{Fneqbis} are written in a way that is manifestly generally covariant, i.e.~invariant under arbitrary changes of the coordinates $x^i\to x^{'i}=f^i(x^j)$. In particular, one can take advantage of this invariance to try and solve these equations in whatever coordinates make our task easiest.

\subsection{Solving the equations\label{Fnsolsec}}

To solve the previous system of equations one has two options: either one goes to normal coordinates $\xi^i$ in which $\del_i \ell^2=2\xi^i$ is particularly simple, but one has to transform all the $g^{ij}_{(r)}$ and $a^i_{(r)}$ into the corresponding expressions in these normal coordinates,
using the relations analogous to \eqref{cordinatetransf} with $\frac{\del x^{'i}}{\del x^k} = \frac{\del \xi^i}{\del x^k}$,
or one keeps the original coordinates and uses the expressions for $\del_i\ell^2$ in these coordinates as given in the appendix \ref{app2}. Let us first keep the original coordinates $x^i$.

The geodesic distance  $\ell^2(x,y)$ between the points $x$ and $y$ can be given as an expansion in $\e^i=x^i-y^i$. One has
\be\label{ellgeneq}
\ell^2(x,y)=g_{ij}(y)\e^i\e^j + {\cal O}(\e^3)\  ,
\quad \e^i=x^i-y^i \ .
\ee
It trivially follow that
\be\label{ellder1}
\del_i \ell^2(x,y)\equiv \frac{\del \ell^2(x,y)}{\del x^i}=2 \,g_{ik}(y)\e^k +\ {\cal O}(\e^2) \ .
\ee
In the appendix \ref{app2} we have given these expansions up to and including terms of order $\e^4$ for $\ell^2(x,y)$ and of order $\e^3$ for its derivative.

Now, $K_0(t,x,y)$ vanishes exponentially unless $\ell^2(x,y)$ is not much larger than a few times $t$, i.e. we may consider that $\ell^2(x,y)$ is of order $t$ and thus $\e=x-y$ is of order $\sqrt{t}$. Thus, the small $t$ expansion is at the same time a small $\e=x-y$ expansion. If we only are interested in the leading small-$t$ behaviour of $K(t,x,y)$, and only keep $F_0$ in $F$, then, consistenly, we must also drop all terms in $F_0(x,y)$ that are of order 1 or higher in $\e$. This means we should also replace $F_0(x,y)\to F_0(y,y)=1$. Of course, this is consistent with \eqref{F0eq} since  $g^{ij}\del_i\del_j\ell^2=2d + {\cal O}(\e)$ and to lowest order \eqref{F0eq} just states $4 \e^i \del_i F_0 = 0$, i.e.~$F_0$ is constant. Thus
\be\label{Klowestorder}
K(t,x,y) = K_0(t,x,y) \big[ 1 + {\cal O}(t) \big]\ ,
\ee
as expected, of course.

\subsubsection{The next-to-leading order corrections $\sim t\sim (x-y)^2$}\label{Fnsolsec1}

We will now work out the first correction, i.e.~the ${\cal O}(t)$-terms. As just discussed, this will involve the term $t F_1(y,y)$ as well as the development of $F_0(x,y)$ up to order $(\e)^2=(x-y)^2$. As already said, we will do this directly using the original coordinates $x^i$ and the forms of $g_{ij}$, $a^i$, $g_{(1)}^{ij}$, etc, as they appear in the Fokker-Planck equation.

We begin by determining $F_0(x,y)$ to this order. We let
\be\label{F0exp}
F_0(x,y)=1+f_i(y)\e^i + f_{ij}(y) \e^i\e^j + {\cal O}(\e^3) \ .
\ee
Let us discuss the various terms in \eqref{F0eqbis}. As shown in appendix \ref{app2}, $\D \ell^2-2d$ is of order $\e^2$, as is obviously the term $g^{ij}_{(1)}\del_i\ell^2\del_j\ell^2$. It is only the term involving the drift vector $a^i$ that gives a first order term in $\e$.  Developping this term to order $\e^2$, using also \eqref{deliell} gives
\be\label{aiexp}
a^i(x) \del_i\ell^2(x,y)= 2g_{ij}a^i\e^j + \big(2g_{il}\del_k a^i+\frac{3}{2} \del_{(k}g_{il)}a^i\big)\e^k\e^l +{\cal O}(\e^3) \ ,
\ee
where all quantities on the right-hand side are evaluated at $y$. From \eqref{Laplell} we know that
\be\label{Laplell2}
-\D_x \ell^2(x,y)+2d=\frac{2}{3} {\cal R}_{kl}\e^k\e^l +{\cal O}(\e^3) \ ,
\ee
where ${\cal R}_{kl}$ is the Ricci curvature tensor of the metric $g_{ij}$. 
Alltogether, the order $\e^2$ terms on the right-hand side of \eqref{F0eqbis} multiplying $F_0$ are $\g_{ij}\e^i\e^j$ with
\be\label{gammadef}
\g_{ij}=g_{ik}\,g_{(1)}^{kl}\,g_{lj}+\frac{2}{3}{\cal R}_{ij}
-2g_{l(i}\del_{j)} a^l-\frac{3}{2} \del_{(k}g_{ij)}a^k
\ee
Finally, we need the term appearing on the left-hand side of \eqref{F0eqbis}. From \eqref{gijdeliell} we see that
\be\label{gijdeliell2}
g^{ij}(x)\frac{\del\ell^2(x,y)}{\del x^j}
=2\e^i-\G^i_{kl}\e^k\e^l +{\cal O}(\e^3) \ .
\ee
Inserting all these expansions into \eqref{F0eqbis} we get
\be\label{F0eqexp}
4f_i\e^i +8f_{ij}\e^i\e^j -2\G^k_{ij}f_k\e^i\e^j +{\cal O}(\e^3)
=-2a_i\e^i -2 a_if_j\e^i\e^j +\g_{ij}\e^i\e^j +{\cal O}(\e^3) \ ,
\ee
where we set, as usual, $a_i=g_{ik}a^k$.
Identifying the various orders in $\e$ yields
\be\label{fisol}
f_i=-\frac{1}{2}g_{ij}a^j \equiv -\frac{a_i}{2} \ ,
\ee
and then
\ba\label{fijsol}
f_{ij}&=&\frac{1}{8}\big( \g_{ij}+a_i a_j -\G^m_{ij}a_m\big)
\nonumber\\
&=&\frac{1}{12}{\cal R}_{ij}+\frac{1}{8}g_{ik}\,g_{(1)}^{kl}\,g_{lj}-\frac{1}{8}(\del_i a_j+\del_j a_i) +\frac{1}{8}a_i a_j\ .
\ea
Thus, we arrive at
\be\label{F0sol}
F_0(x,y)=1-\frac{1}{2}a_i\e^i+ \Big(\frac{1}{8}a_ia_j+\frac{1}{12}{\cal R}_{ij}+\frac{1}{8}g_{ik}\,g_{(1)}^{kl}\,g_{lj}-\frac{1}{8}(\del_i a_j+\del_j a_i)\Big)\e^i\e^j+{\cal O}(\e^3) \quad , \quad \e^i=x^i-y^i \ ,
\ee
with all terms on the right-hand-side evaluated at $y$. (Also, all quantities in this expression only involve the initial metric and drift coefficients, as well as their time derivatives at the initial time $t_0$).
Of course, the terms involving only the $a_i$ without derivatives combine into
$e^{-a_i\e^i/2}$ as they should.\footnote{ 
The careful reader might worry that this expression for $F_0$ does not seem to be a scalar since $(\del_i a_j+\del_j a_i)$ does not involve the covariant derivatives of the vector $a_i$ but only ordinary derivatives. However, $\e^i$ being a coordinate difference is not a vector either. Using \eqref{epseta} we may express $\e^i$ in terms of $\eta^i$ which is a true vector: $\e^i=\eta^i-\frac{1}{2} \G^i_{jk}\eta^j\eta^k +{\cal O}(\e^3)$. Thus $-\frac{1}{2}a_i\e^i=  -\frac{1}{2}a_i\eta^i+\frac{1}{4} a_l\G^l_{ij}\eta^i\eta^j+{\cal O}(\eta^3)$, providing just the term needed to turn $-\frac{1}{8}(\del_i a_j+\del_j a_i)\e^i\e^j$ into $-\frac{1}{8}(\nabla_i a_j+\nabla_j a_i)\eta^i\eta^j$, up to terms of order $\eta^3$.}

Next, we insert this result into the equation for $F_1$. As explained above, $t$ is of the same order as $\e^2$ and since $F_1$ is multiplied by $t$, we will only determine $F_1$ to order 0 in $\e$, i.e. we only need $F_1(y,y)$. To this order, \eqref{Fneqbis}  for $n=0$ becomes:
\be\label{F1eq}
F_1=\Big[-\frac{1}{4}(\D_{(1)}\ell^2)+ \big( \D+a^i \del_i +c\big)\Big] F_0 +{\cal O}(\e) \ .
\ee
Now, using eq.~\eqref{deldeliell} we have $\D_{(1)}\ell^2=g^{ij}_{(1)}\del_i\del_j \ell^2+{\cal O}(\e)=2 g^{ij}_{(1)} g_{ij}+{\cal O}(\e)$.
Recalling also  the form \eqref{Laplaciansr} of $\D$, we find from \eqref{F0sol} that
\ba\label{DonF0}
\big( \D+a^i \del_i +c\big) F_0
&=&2g^{ij} f_{ij} +(a^i-g^{rs}\G^i_{rs})f_i +c+{\cal O}(\e)
\nonumber\\
&=&\frac{1}{6} {\cal R} +\frac{1}{4} g^{ij}_{(1)} g_{ij}-\frac{1}{2}\nabla_i a^i -\frac{1}{4} a_i a^i+c +{\cal O}(\e)\ .
\ea
Inserting these results into \eqref{F1eq} we finally get
\be\label{F1eq2}
F_1\equiv F_1(y,y)=\frac{1}{6} {\cal R} -\frac{1}{2}\nabla_i a^i -\frac{1}{4} a_i a^i +c -\frac{1}{4} g^{ij}_{(1)} g_{ij} +{\cal O}(\e) \ .
\ee
Of course, for $a^i=c=g^{ij}_{(1)}=0$, equations  \eqref{F0sol}  and \eqref{F1eq2} reduce to the well-known results for the heat kernel of the scalar Laplace operator on a manifold with metric $g_{ij}$.

To summarize, up to and including terms of order $t\sim \ell^2(x,y)=g_{ij}\e^i\e^j + {\cal O}(\e^3)$, the Fokker-Planck kernel $K(t,x,y)$ is given,  by
\ba\label{summary}
K(t,x,y)&=& K_0(t,x,y) \Big(F_0(x,y) + t F_1(y,y) + \ldots \Big)  \nonumber\\
&=& \frac{1}{(4\pi t)^{d/2}} \exp\Big( -\frac{\ell^2(x,y)}{4t}\Big) \Big(F_0(x,y) + t F_1(y,y) + \ldots \Big)  \ ,
\ea
with $F_0(x,y)$ and $F_1(y,y)$ given by \eqref{F0sol} and \eqref{F1eq2}, and $\ell^2(x,y)$ by eq.~\eqref{geodistcoord}, namely
\be\label{geodistcoordmain}
\ell^2(x,y)=g_{ij}\e^i\e^j +\frac{1}{2} \del_k g_{ij} \e^k\e^i\e^j 
+  \big( \frac{1}{6} \del_i\del_j g_{lk} -\frac{1}{12} g_{nm}\G^n_{ij}\G^m_{kl} \big)\e^i\e^j\e^k\e^l +\ldots \ ,
\ee
The only place where the time-dependence of the diffusion matrix $G^{ij}$ shows up at this order is through the terms $g_{ik}g^{kl}_{(1)} g_{lj}$ in $F_0$ and $g_{(1)}^{ij} g_{ij}$ in $F_1$. Any time dependence of the drift coefficients $B^i$ only enter the equations at the next order. With quite some patience, these higher orders can be worked out along the same lines, as we partly show in the next sub-section.

\newpage
\subsubsection{The next-to-next-to-leading order corrections $\sim t^2\sim t (x-y)^2\sim (x-y)^4$}\label{Fnsolsec2}

At the next order, we should keep the terms $\sim (x-y)^4\sim \e^4$ in $F_0(x,y)$, the terms $\sim(x-y)^2\sim \e^2$ in $t F_1(x,y)$, while we only need $t^2 F_2(y,y)$. Compared to the previous computation, the number of terms present is considerably larger, so we will only present our results for the special cases
\be\label{a=c=0}
a^i=c=0\ .
\ee
The time dependence then only shows up through the tensors $g_{(1)}^{ij}$ and $g_{(2)}^{ij}$. The higher $g_{(r)}^{ij}$, $r\ge 3$ will not enter the formula to this order. As before, we may lower the indices using the metric $g_{ij}$~:
\be\label{grlower}
g_{(r)j}^{\ i}=g_{(r)}^{il}\,g_{lj} \quad , \quad g_{ij}^{(r)}=g_{ik}\,g_{(r)}^{kl}\,g_{lj} \ .
\ee

Moreover, to simplify things further, we will use (Riemann) normal coordinates called $\xi^i$, centered around the point corresponding to $y$ (i.e.~$y$ corresponds to $\xi=0$). Some properties of these normal coordinates are recalled in the appendix \ref{app2}. Of course, the normal coordinates are defined with respect to the metric $g_{ij}$.
In general, the initial coordinates of the Fokker-Planck equation will not be normal coordinates and the metric $g_{ij}$, as well as the tensors $g_{(1)}^{ij}$ and $g_{(2)}^{ij}$ will not be given directly in this normal coordinate system. However, it is often not too difficult to realise the transformation and obtain these quantities in the normal coordinate system. For the rest of this sub-section, $g_{ij}$, $g_{(1)}^{ij}$ and $g_{(2)}^{ij}$ denote the components of these tensors in the normal coordinate system.

The important relations satisfied by the normal coordinates $\xi^i$ are (cf Appendix \ref{app2})
\ba\label{normalproperties}
&&\ell^2(x,y)\equiv \ell^2(\xi,0)=\sum_k \xi^k\xi^k \ , \quad 
g_{ij}(\xi)=\dd_{ij}-\frac{1}{3} R_{ikjl}(0)\xi^k\xi^l + {\cal O}(\xi^3)\ , \nonumber\\
&&g_{ij}(\xi) \xi^j =g^{ij}(\xi) \xi^j = \dd_{ij}\xi^j\equiv \xi^i \quad \Rightarrow\quad 
g^{ij}(\xi)\, \del_j \ell^2=2\xi^i \ .
\ea
We will need the expression of $\D\ell^2$, which is worked out in the appendix \ref{app3} to second order in $x-y$ for arbitrary coordinates, and can be found in \cite{BF} to fourth order in $\xi$ for normal coordinates:
\be\label{Laplellmain}
-\D \ell^2 + 2 d = \frac{2}{3}\cR_{kl}(0)\xi^k\xi^l +\frac{1}{2} \cR_{kl;m}(0)\xi^k\xi^l\xi^m +E_{klmn}(0)\xi^k\xi^l\xi^m\xi^n  +{\cal O}(\xi^5)\ ,
\ee
where $\cR_{kl;m}=\nabla_m \cR_{kl}$ indicates a covariant derivative of the Ricci tensor, and
\be\label{E-tensor}
E_{klmn}=\frac{1}{5}\cR_{kl;mn}+\frac{2}{45} R^r_{\ kls}R^s_{\ mnr} \ ,
\ee
where $\cR_{kl;mn}\equiv \nabla_n\nabla_m \cR_{kl}$ indicates a second covariant derivative of the Ricci tensor.

We note that $g_{(1)}^{ij} \del_i\ell^2 \del_j\ell^2= g^{(1)}_{mn}\, g^{mi}\del_i\ell^2 g^{nj}\del_j\ell^2=4 g^{(1)}_{mn}\, \xi^m\xi^n$, so that
the equation \eqref{F0eqbis} for $F_0$  reads 
\be\label{F0eq3}
4\xi^j\, \del_j F_0 = \Big(g^{(1)}_{kl}(\xi)\, \xi^k \xi^l -(\D\ell^2) +2d  \Big) F_0  \ ,
\ee
Using \eqref{Laplellmain}, and expanding $g^{(1)}_{ij}(\xi)$ in a series around $\xi=0$, this becomes
\be\label{F0eq4}
4\xi^j\, \del_j F_0 = \Bigg[
\Big(g^{(1)}_{kl} + \frac{2}{3}\cR_{kl}\Big)\xi^k \xi^l
+\Big(\del_m g^{(1)}_{kl}+\frac{1}{2}\cR_{kl;m}\Big)\xi^k\xi^l\xi^m 
+\Big( \frac{1}{2} \del_m\del_n  g^{(1)}_{kl}+E_{klmn}\Big)\xi^k\xi^l\xi^m\xi^n +\ldots \Bigg] F_0  \ ,
\ee
where now all quantities $g^{(1)}_{kl}, \ \cR_{kl;m}$, etc in the round brackets on the r.h.s. are to be taken at $\xi=0$. 
Note that the $\G_{ij}^k$ vanish at $\xi=0$, so that first derivatives can be replaced by the corresponding covariant derivatives : $\del_m g^{(1)}_{kl}=g^{(1)}_{kl;m}$.  The rewriting of $\del_m\del_n g^{(1)}_{kl}$ in terms of covariant derivatives generates terms involving  non-vanishing $\del \G$ which can be expressed in terms of the curvature tensor. Here, however, these terms involve $R^i_{\ mnk}g^{(1)}_{il} \xi^k\xi^l\xi^m\xi^m$ which vanishes by the antisymmetry of the curvature tensor. Thus, \eqref{F0eq4} can be  rewritten as
\be\label{F0eq5}
4\xi^j\, \del_j F_0 = \Bigg[
\Big(g^{(1)}_{kl} + \frac{2}{3}\cR_{kl}\Big)\xi^k \xi^l
+\Big(g^{(1)}_{kl;m}+\frac{1}{2}\cR_{kl;m}\Big)\xi^k\xi^l\xi^m 
+\Big( \frac{1}{2}g^{(1)}_{kl;nm}+E_{klmn}\Big)\xi^k\xi^l\xi^m\xi^n +\ldots \Bigg] F_0  \ ,
\ee
This equation is immediately solved as
\ba\label{F0sol4}
F_0(\xi)&&\hskip-6.mm= \exp\Bigg[
\frac{1}{8}\Big(g^{(1)}_{kl} + \frac{2}{3}\cR_{kl}\Big)\xi^k \xi^l
+ \frac{1}{12} \Big(g^{(1)}_{kl;m}+\frac{1}{2}\cR_{kl;m}\Big)\xi^k\xi^l\xi^m 
\nonumber\\
&&\hskip6.mm +\frac{1}{16} \Big( \frac{1}{2} g^{(1)}_{kl;nm}+E_{klmn}\Big)\xi^k\xi^l\xi^m\xi^n +\ldots \Bigg]
\nonumber\\
&&\hskip-6.mm= 1 + \Big(\frac{1}{8}g^{(1)}_{kl} + \frac{1}{12}\cR_{kl}\Big)\xi^k \xi^l
+ \Big( \frac{1}{12}g^{(1)}_{kl;m}+\frac{1}{24}\cR_{kl;m}\Big)\xi^k\xi^l\xi^m 
\nonumber\\
&& + \Big( \frac{1}{32}g^{(1)}_{kl;nm} + \frac{1}{128} g^{(1)}_{kl} g^{(1)}_{mn} 
+\frac{1}{96} g^{(1)}_{kl} \cR_{mn}+\frac{1}{16}E_{klmn}+\frac{1}{288} \cR_{kl}\cR_{mn}\Big)\xi^k\xi^l\xi^m\xi^n +\ldots 
\nonumber\\
&&\hskip-6.mm\equiv 1+\frac{1}{2} \f_{kl} \xi^k\xi^l + \frac{1}{3} \f_{klm} \xi^k\xi^l\xi^m + \frac{1}{4} \f_{klmn} \xi^k\xi^l\xi^m\xi^m + \ldots \ ,
\ea
where the $\f$ are  completely symmetrised coefficients defined by this equation. 

Next, eq \eqref{Fneqbis} for $n=0$  reads
\be\label{F1eqnormal}
\Big( 4\xi^j\, \del_j +\D \ell^2-2d +4 -g^{(1)}_{kl}\, \xi^k\xi^l \Big) F_1
= \Big(g^{(2)}_{kl}\,  \xi^k\xi^l -\D_{(1)} \ell^2 - 4 \xi^k g^{(1)}_{kl}\,   g^{lj}\, \del_j + 4 \D \Big) F_0
\ee
A priori, the $g^{(1)}_{kl}$ and  $g^{(2)}_{kl}$ are to be taken at $\xi$, but since we only want $F_1$ to order $\xi^2$ (and $\del_j F_0$ is of order $\xi$), we may well replace these  $g^{(1)}_{kl}$ and  $g^{(2)}_{kl}$ in this equation by their values at $\xi=0$. Of course, this does not apply to the $g_{(1)}^{ij}$ hidden in $\D_{(1)}$. We have
\be\label{D1l2}
\D_{(1)} \ell^2 = \frac{1}{\sqrt{g}} \del_i \big( g_{(1)}^{ij} \sqrt{g} \,\del_j \ell^2 \big)
=\frac{1}{\sqrt{g}} \del_i \big( g_{(1)k}^{\ i} \sqrt{g}\, g^{kj}\del_j \ell^2 \big)
=\frac{2}{\sqrt{g}} \del_i \big( g_{(1)k}^{\ i} \sqrt{g}\,\xi^k \big) \ .
\ee
Now, using the expression of $g_{rs}$ given in \eqref{normalproperties} we have
 $\del_i \log \sqrt{g}=\frac{1}{2} g^{rs} \del_i g_{rs}=-\frac{1}{3} \cR_{il}\xi^l + {\cal O}(\xi^2)$. It remains to expand $g_{(1)i}^{\ i}(\xi)$ to second order in $\xi$ and  $\del_i g_{(1)k}^{\ i}(\xi)$ to first order in $\xi$, and use the relations \eqref{delnablaT}, to get
\ba\label{D1l2bis}
\D_{(1)} \ell^2&=& 2g_{(1)i}^{\ i}+ 2 \big(\del_k g_{(1)i}^{\ i} + \del_i g_{(1)k}^{\ i}\big)\xi^k
+\Big( \del_k\del_l g_{(1)i}^{\ i} +2\del_i\del_l g_{(1)k}^{\ i}-\frac{2}{3} g_{(1)k}^{\ i} \cR_{il} \Big) \xi^k\xi^l +\ldots 
\nonumber\\
&=&  2g_{(1)i}^{\ i}+ 2 \big(g_{(1)i;k}^{\ i} +g_{(1)k;i}^{\ i}\big)\xi^k
+\Big( g_{(1)i;lk}^{\ i} +2g_{(1)k;il}^{\ i}-\frac{2}{3} g_{(1)}^{ij} R_{iljk} \Big) \xi^k\xi^l +\ldots 
\ ,
\ea
where, again, all quantities on the r.h.s are to be taken at $\xi=0$.  Using again \eqref{Laplellmain} up to order $\xi^2$ and   $\del_i g^{ij}(\xi)=-\frac{1}{3} \cR_{il}\xi^l + {\cal O}(\xi^2)$ we have
\ba\label{LaplFoxi2}
\hskip-1.cm\D F_0&=& g^{ij}(\xi) \del_i\del_j F_0 + \big(\del_i g^{ij}(\xi) + \del_i \log \sqrt{g}(\xi) \big) \del_j F_0
\nonumber\\
&=& \big(g^{ij}+\frac{1}{3} R^{i\ j}_{\ k \ l}\xi^k\xi^l+\ldots\big) \del_i\del_j F_0+\big( -\frac{2}{3} \cR^j_{\ k}\xi^k + \ldots)\del_j F_0 \nonumber\\
&=& g^{ij}\big(\f_{ij} +2\f_{ijm}\xi^m +3\f_{ijmn}\xi^m\xi^n\big) +\frac{1}{3} R^{i\ j}_{\ k \ l}\, \f_{ij}\xi^k\xi^l
-\frac{2}{3} \cR^j_{\ k}\xi^k \, \f_{jl}\xi^l +\ldots\ ,
\ea
where, again, all quantities with no $\xi$-dependence indicated are to be taken at $\xi=0$. It remains to insert this relation, as well as \eqref{D1l2bis} and \eqref{F0sol4} into the differential equation \eqref{F1eqnormal}. Although a bit lengthy, it is then completely straightforward to solve for $F_1$ in an expansion
\be\label{F1expnormal}
F_1(\xi)=\vf +\vf_k \xi^k + \frac{1}{2}\vf_{kl}\xi^k\xi^l +\ldots 
\ee
The result is\footnote{
To write the result in this form one has to use  $2\cR_{k\ ;j}^{\ j}=\cR_{;k}$ which follows from the Bianchi identity for the curvature tensor.}
\ba\label{F1expnormalter}
\hskip-0.7cmF_1(\xi)&\hskip-2.mm=\hskip-2.mm&\frac{1}{6} \cR - \frac{1}{4} g_{(1)i}^i + \Big( \frac{1}{12} \cR_{;k} 
-\frac{1}{6} g_{(1)i;k}^i  -\frac{1}{12}  g_{(1)k;i}^i \Big) \xi^k
\nonumber\\
&&\hskip-3.mm+\ \Bigg( \frac{1}{2}\vf_{kl}^{(0)}-\frac{1}{24}g_{(1)}^{ij}R_{ikjl}
+\frac{1}{48}g_{(1)k}^i\cR_{il} + \frac{1}{48}g_{(1)l}^i\cR_{ik}+\frac{1}{48} g^{(1)}_{kl} \cR -\frac{1}{48} g_{(1)i}^i \cR_{kl} 
\nonumber\\
&&\hskip-0.mm + \frac{1}{48}g^{ij}g^{(1)}_{kl;ij}-\frac{1}{16} g_{(1)i;kl}^i
-\frac{1}{24} g_{(1)k;li}^i-\frac{1}{24} g_{(1)l;ki}^i
 -\frac{1}{32} g_{(1)i}^i g^{(1)}_{kl} -\frac{1}{16} g_{(1)k}^i g^{(1)}_{il} +\frac{1}{12} g^{(2)}_{kl}\Bigg) \xi^k\xi^l 
 \nonumber\\
&&\hskip-2.mm +\ldots \ .
\ea
%
%
where $\frac{1}{2} \vf_{kl}^{(0)}\xi^k\xi^l$ are the $\xi^2$-terms already present for $g_{(1)}=g_{(2)}=0$~:
\be\label{phizero}
\frac{1}{2} \vf_{kl}^{(0)}=\frac{1}{90} g^{ij}R^r_{\ (ij|m}R^m_{\ \ |kl)r} +\frac{1}{54}\cR^{ij} R_{ikjl}+\frac{1}{20}g^{ij}\cR_{(ij;kl)}
+\frac{1}{72}\cR \cR_{kl}-\frac{1}{36}\cR_{ik}\cR^i_{\ l} \ .
\ee
Of course, they coincide with those given e.g.~in \cite{BF}. 

Finally, we work out $F_2$ which we only need for $\xi=0$. Thus, dropping all terms involving at least one $\xi$, eq \eqref{Fneqbis} for $n=1$  reads
\be\label{F2eqnormal}
8 F_2
= \Big( 4 \D -\D_{(1)} \ell^2 \Big) F_1 +  \Big( 4 \D_{(1)}-\D_{(2)} \ell^2  \Big) F_0 \ .
\ee
Recall from  \eqref{D1l2bis} that
$\D_{(1)} \ell^2= 2g_{(1)i}^{\ i} +{\cal O}(\xi)$, and in exactly the same way also $\D_{(2)} \ell^2= 2g_{(2)i}^{\ i} +{\cal O}(\xi)$. Next, on easily sees that $\D F_1=g^{ij}\del_i\del_j F_1+{\cal O}(\xi)=g^{ij}\vf_{ij}+{\cal O}(\xi)\equiv \vf^i_{\ i}+{\cal O}(\xi)$. Similarly, $\D_{(1)} F_0=g_{(1)}^{ij} \del_i\del_j F_0 +{\cal O}(\xi)= g_{(1)}^{ij} \f_{ij}+{\cal O}(\xi)$, so that
\be\label{F2eqnormalbis}
F_2(0)=\frac{1}{2}\vf^i_{\ i} -\frac{1}{4} g_{(1)i}^i \,\vf+\frac{1}{2} g_{(1)}^{ij}\f_{ij}-\frac{1}{4} g_{(2)i}^i \ .
\ee
Inserting the expressions for $\vf,\ \vf^i_{\ i}$ and $\f_{ij}$ read from \eqref{F1expnormalter} and \eqref{F0sol4}, we find 
\be\label{F2eqnormalter}
F_2(0)=\frac{1}{2}\vf_{i}^{(0)i}-\frac{1}{24} g_{(1)i}^i \cR+\frac{1}{12} g_{(1)}^{ij}\cR_{ij}
+\frac{1}{32} g_{(1)i}^i  g_{(1)j}^j +\frac{1}{16} g_{(1)}^{ij}g^{(1)}_{ij}-\frac{1}{6} g_{(2)i}^i
-\frac{1}{24} g^{ij}g_{(1)k;ij}^k -\frac{1}{12} g_{(1);ij}^{ij} \ ,
\ee
where $\frac{1}{2}\vf_{i}^{(0)i}$ can be read from \eqref{phizero}.

\section{Examples\label{examples}}

To illustrate the formulae of the previous section, we will present some trivial and some less trivial examples.
The trivial examples can be solved by some ``reparametrisation-trick" and provide a non-trivial consistency check of our above results.

\subsection{Constant coefficients}

Suppose that the coefficents $G^{ij}$, $B^i$ and $C$ depend neither on $x$ nor on $t$, i.e.
\be\label{constcoeff}
G^{ij}=g^{ij} \ , \quad B^i=b^i=a^i \ , \quad C=c  \ , \quad
\del_k g^{ij}=\del_k a^i=\del_k c=0 \ .
\ee
Since the metric is constant the geodesics are just affine functions of the coordinates and the geodesic length is $\ell^2(x,y)=g_{ij}(x^i-y^i)(x^j-y^j)$. Of course, the curvature vanishes.
By performing the change of variables\footnote{
This is not a coordinate transformation in the sense discussed before since it mixes the coordinates $x^i$ and the time $t$.}
\be\label{xtmix}
t'=t\quad , \quad x^{'i}=x^i + a^i t
\quad \Rightarrow \quad
\frac{\del}{\del t}=\frac{\del}{\del t'} + a^i \del'_i \quad , \quad \del_i=\del'_i \ .
\ee
the Fokker-Planck equation for $K$ becomes
\be\label{FPprime}
\frac{\del}{\del t'} K=\big( g^{ij} \del'_i\del'_j + c \big) K \ ,
\ee
without a drift term. Its solution is
\ba\label{Kconstant}
K(t,x,y)&=&\frac{1}{(4\pi t')^{d/2}} \exp\left(- \frac{g_{ij}(x^{'i}-y^i)(x^{'j}-y^j)}{4t'} +c\, t'\right)
\nonumber\\
&=&\frac{1}{(4\pi t)^{d/2}} \exp\left(- \frac{g_{ij}(x^i-y^i+a^i t)(x^j-y^j+a^j t)}{4t} +c\, t\right)
\ea
Expanding this in powers of $t$ to next-to-leading order, we get
\be\label{Kconstantexp}
K(t,x,y)
=\frac{1}{(4\pi t)^{d/2}} \exp\left(- \frac{g_{ij}(x^i-y^i)(x^j-y^j)}{4t}\right)
\exp\left(-\frac{1}{2}a_i(x-y)^i\right) \Big[ 1 + t\big( c-\frac{1}{4}a_i a^i \big) +{\cal O}(t^2)\Big] \ ,
\ee
perfectly consistent with the previously  results \eqref{F0sol} for $F_0$ and \eqref{F1eq2} for $F_1$.

\subsection{A trivial time-dependence providing a non-trivial check}

Suppose the coefficients $G^{ij},\ A^i$ and $C$ depend on time only through a common factor which we write as $\dot f(t)=\frac{\del f}{\del t}(t)$. (As already mentioned in the introduction, the examples discussed in \cite{WU} fall in this class.) Without loss of generality, we assume $f(0)=0$ and $\dot f(0)=1$,  and obviously also $\dot f(t)\ne 0\  \forall\  t$. Thus
\be\label{trivialtcoeffs}
G^{ij}(t,x)=\dot f(t) g^{ij}(x) \ , \quad
A^i(t,x)=\dot f(t) a^i(x) \ , \quad
C(t,x)=\dot f(t) c(x) \ .
\ee
Then changing the time variable from $t$ to 
\be\label{ttprime}
t'=f(t) \quad \Rightarrow \quad \frac{\del}{\del t}=\dot f(t) \frac{\del}{\del t'}
\ee
the Fokker-Planck equation for $K(t,x,y)$  reads\footnote{
Note that by \eqref{VG}, for all $t$ we have $e^{-2V(t,x)}=\det G^{ij}(0,x)=\det (\dot f(0) g^{ij}(x))=\det g^{ij}(x)$ so that $e^V=\sqrt{g}$.
}
\be\label{FPtprime}
\frac{\del}{\del t'} K = \left[ \frac{1}{\sqrt{g}}\del_i ( \sqrt{g}g^{ij}(x) \del_j) + a^i(x) \del_i + c(x) \right] K
\ee
This is a standard Fokker-Planck equation in time $t'$ with time-independent coefficients. Thus if $\wt K(t',x,y)$ is a solution of \eqref{FPtprime} then $K(t,x,y)=\wt K(t',x,y)$ will solve our Fokker-Planck equation with the time-dependent coefficients. Obviously, the small $t'$ expansion of \eqref{FPtprime}  is given by the standard  expansion \cite{Vassil,BF}, as also given in the previous section but without the terms involving $g^{ij}_{(r)}, \ a^i_{(r)},\ c_{(r)}$ for $r\ge 1$. We will now show how the change of time variable generates the extra terms we have determined in the previous section, thus providing a highly non-trivial consistency check.  At the next-to-leading order, the time-dependence of the coefficients only will show up through $g^{ij}_{(1)}$, while the $a^i_{(1)}$ and $c_{(1)}$ would only show up at the next-to-next-to-leading order where we had set them to zero to simplify our formulae. 

Thus, to check our previous results we now concentrate on the case   $A=C=0$. Using normal coordinates we have
\ba\label{Ktprime}
K(t,x,y)&\equiv& \wt K(t',x,y)
\nonumber\\
&=&\frac{1}{(4\pi t')^{d/2}} \exp\left(- \frac{\ell^2(\xi,0)}{4t'} \right)
\Bigg\{\Big[1+\frac{1}{12}{\cal R}_{kl}\xi^k\xi^l+\frac{1}{24}\cR_{kl;m}\xi^k\xi^l\xi^m
\nonumber\\
&&\hskip5.cm + \Big(\frac{1}{16} E_{klmn} +\frac{1}{288} \cR_{kl}\cR_{mn}\Big) \xi^k\xi^l\xi^m\xi^n \Big]
\nonumber\\
&&\hskip5.cm +  t' \Big[ \frac{1}{6} {\cal R}+\frac{1}{12} \cR_{;k}\xi^k + \frac{1}{2}\vf^{(0)}_{kl}\xi^k\xi^l\Big]
+ (t')^2 \vf^{(0)k}_k + \ldots \Bigg\} \ .
\nonumber\\
\ea
Our assumptions about the function $f(t)$ imply that
\be\label{tt'rel2}
t'=f(t)= t+\frac{\ddot f(0)}{2} t^2 +\frac{\dddot f(0)}{6} t^3 + {\cal O}(t^4)
\equiv t+\frac{\a}{2}t^2 + \frac{\b}{3} t^3  + {\cal O}(t^4)\ ,
\ee
so that $\dot f(t)=1+\a t+\b t^2 +{\cal O}(t^3)$ and
\be\label{grarcr}
g^{ij}_{(1)}(x)=\a\, g^{ij}(x) \ , \  g^{ij}_{(2)}(x)=\b\, g^{ij}(x)  \ .
\ee
In particular, (recall that $d=g_i^i$ is the dimension of space)
\be\label{alphabetaid}
g_{(1)i}^i=d\a \ , \quad g_{(2)i}^i= d \b \ , \quad g_{(1)}^{ij} g^{(1)}_{ij} =d \a^2 \ .
\ee
If we now insert \eqref{tt'rel2} into \eqref{Ktprime}, expand for small $t$ and use \eqref{grarcr} we should recover our expansion from the previous section. 

To see this, first note that 
\ba\label{alphaexp}
\frac{1}{(4\pi t')^{d/2}}&=&\frac{1}{(4\pi t)^{d/2}}\Big[1-\frac{d}{4}\a t +\frac{d^2}{32}\a^2 t^2 +\frac{d}{16}\a^2 t^2 -\frac{d}{6}\b t ^2  +{\cal O}(t^3)\Big] 
\nonumber\\
&=&\frac{1}{(4\pi t)^{d/2}}\Big[1-\frac{t}{4}g_{(1)i}^i + t^2 \Big(  \frac{1}{32}g_{(1)i}^i g_{(1)j}^j  +\frac{1}{16}g_{(1)}^{ij} g^{(1)}_{ij} -\frac{1}{6}g_{(2)i}^i  \Big) +{\cal O}(t^3)\Big] \ .
\ea
Next, recall that we must consider $\ell^2(\xi,0)=g_{kl}\xi^k\xi^l$ as being of order $t$, so that
\ba\label{alphaexp2}
\hskip-2.mm\exp\left(- \frac{\ell^2}{4t'} \right)&\hskip-2.mm=\hskip-2.mm& \exp\left(- \frac{\ell^2}{4t} \right)
\Big[ 1 + \frac{\a}{8}\ell^2 +\frac{\b}{12}\ell^2 t -\frac{\a^2}{16}\ell^2 t +\frac{\a^2}{128} (\ell^2)^2  + \ldots  \Big]
\nonumber\\
&\hskip-2.mm=\hskip-2.mm&\exp\left(- \frac{\ell^2}{4t} \right)
\Big[ 1 +\frac{1}{8} g^{(1)}_{kl}\xi^k\xi^l +\frac{1}{128} g^{(1)}_{kl} g^{(1)}_{mn}\xi^k \xi^l \xi^m\xi^n
+ t \Big( \frac{1}{12} g^{(2)}_{kl} -\frac{1}{16} g^{(1)}_{kj}g^{(1)j}_l \Big) \xi^k\xi^l +\ldots
\Big]
\nonumber\\
\ea
Inserting \eqref{alphaexp} and \eqref{alphaexp2} into \eqref{Ktprime}, and writing $t'=t+\frac{\a}{2}t^2 +\ldots$,
we see that the time-dependence of the coefficients in the Fokker-Planck equation generates  additional contributions to $F_0$,  $F_1$ and $F_2$. To the order we work, we find
\ba\label{F0-F2}
F_0&=&1 + \Big(\frac{1}{8}g^{(1)}_{kl} + \frac{1}{12}\cR_{kl}\Big)\xi^k \xi^l
+ \frac{1}{24}\cR_{kl;m} \xi^k\xi^l\xi^m 
\nonumber\\
&& + \Big(  \frac{1}{128} g^{(1)}_{kl} g^{(1)}_{mn} 
+\frac{1}{96} g^{(1)}_{kl} \cR_{mn}+\frac{1}{16}E_{klmn}+\frac{1}{288} \cR_{kl}\cR_{mn}\Big)\xi^k\xi^l\xi^m\xi^n +\ldots \ ,
\nonumber\\
\nonumber\\
F_1&=&\frac{1}{6} \cR - \frac{1}{4} g_{(1)i}^i + \frac{1}{12} \cR_{;k}  \xi^k
\nonumber\\
&&\hskip-3.mm+\ \Bigg( \frac{1}{2}\vf_{kl}^{(0)}-\frac{1}{48} g_{(1)i}^i \cR_{kl} 
+\frac{1}{48} g^{(1)}_{kl} \cR
-\frac{1}{32} g_{(1)i}^i g^{(1)}_{kl} -\frac{1}{16} g_{(1)k}^i g^{(1)}_{il} +\frac{1}{12} g^{(2)}_{kl}\Bigg) \xi^k\xi^l 
 +\ldots \ ,
\nonumber\\
\nonumber\\
F_2&=&\frac{1}{2}\vf_{i}^{(0)i}-\frac{1}{24} g_{(1)i}^i \cR+\frac{1}{12} g_{(1)}^{ij}\cR_{ij}
+\frac{1}{32} g_{(1)i}^i  g_{(1)j}^j +\frac{1}{16} g_{(1)}^{ij}g^{(1)}_{ij}-\frac{1}{6} g_{(2)i}^i +\ldots \ .
\ea
Note that at present, all terms involving covariant derivatives of $g^{(1)}_{ij}$ vanish since $g^{(1)}_{ij;k}=\a g_{ij;k}=0$ (for all $\xi$). Furthermore, $g_{(1)}^{ij} R_{kilj}=\a \cR_{kl}=g_{(1)k}^i\cR_{ij}$. Thus we see that
the $F_0,\ F_1$ and $F_2$ as given in \eqref{F0-F2} exactly correspond to the general $F_0,\ F_1$ and $F_2$ worked out above in \eqref{F0sol4}, \eqref{F1expnormalter} and \eqref {F2eqnormalter}. The study of this trivial example has provided a non-trivial check of our general results!


\subsection{A non-trivial example}

In this subsection we present an example that is neither trivial, nor particularly simple. The goal is to illustrate on this example that it is completely straightforward to explicitly work out all expressions appearing in our expansion \eqref{summary} of the Fokker-Planck kernel.

Consider a distribution $\r(t,u,v)$ depending on 2 coordinates $x^1\equiv u$ and $x^2\equiv v$, with Fokker-Planck equation
\be\label{FPex3}
\frac{\del}{\del t}\r=\Big(1+ \frac{\g\,t}{1+t^2}\, \frac{u^2 v^2}{1+u^2 v^2}\Big)\del_u^2 \,\r 
+\Big( 1+\frac{u^2}{1+u^2}\Big)\del_v^2\,\r -\frac{u}{1+u^2}\del_u\,\r \ .
\ee
From this we read $g^{11}=1,\ g^{22}=\frac{1+2u^2}{1+u^2},\ g^{12}=g^{21}=0$, $b^1=-\frac{u}{1+u^2},\ b^2=0,\ c=0$, as well as $g^{11}_{(1)}= \g\, \frac{u^2 v^2}{1+u^2 v^2},\ g^{22}_{(1)}=g^{12}_{(1)}=g^{21}_{(1)}=0$. The metric $g_{ij}$ then is
\be\label{metricex}
g_{11}=1 \ , \quad g_{22}=\frac{1+u^2}{1+2u^2} \ , \quad g_{12}=g_{21}=0 \ .
\ee
\vskip-2.mm
\noindent
Note that for $u\to\infty$, $g_{22}\to \frac{1}{2}$ and the metric becomes flat. It is straightforward to compute the Christoffel symbols and curvature tensor. To save some writing, it is convenient to introduce the notation $U_1=1+u^2$ and $U_2=1+2u^2$. Then
\be\label{Christoffelex}
\G^1_{22}=\frac{u}{U_2^2} \ , \quad \G^2_{12}=\G^2_{21}=-\frac{u}{U_1 U_2} \ , \quad
\G^1_{12}=\G^1_{21}=\G^1_{11}=\G^2_{11}=\G^2_{22}=0\ .
\ee 
(Although the $\G$ vanish at $u=0$, the present coordinates are {\it not} normal coordinates around the line $u=0$. This will be clear from the expressioon of the geodesic distance $\ell^2$ given below.)
Due to the antisymmetry properties of the curvature tensor, in 2 dimensions there is only one independent component which we may take to be $R_{1212}$. Then
\be\label{curvatureex}
R_{1212}=\frac{1-4u^2-6u^4}{U_1 U_2^3} \ ,\quad
{\cal R}_{ij}=\frac{1-4u^2-6u^4}{U_1^2 U_2^2}\ g_{ij}\ .
\ee
\vskip-1.mm
\noindent
The geodesic length between a point with coordinates $(u,v)$ and another point with coordinates $(u',v')\equiv(u+\D u, v+\D v)$ can then be obtained from \eqref{geodistcoord} to the order we consider (we write $\D u^2$ and $\D v^2$ instead of $(\D u)^2$ and $(\D v)^2$):
\be\label{elluv}
\ell^2\big((u', v'),(u,v)\big)=\D u^2 +\frac{U_1}{U_2}\D v^2 -\frac{u}{U_2^2}\D u \D v^2-\frac{1-6u^2}{3 U_2^3}\D u^2 \D v^2 -\frac{u^2}{12 U_2^4} \D v^4 -\frac{u^2}{12 U_1 U_2^3}\D u^2 \D v^2 +\ldots
\ee
\vskip-3.mm
\noindent
Next (cf \eqref{abrelGamma}), 
\be\label{aresults}
a^1=b^1+g^{22}\G^1_{22}=-\frac{2 u^3}{U_1 U_2} \ , \quad a^2=0 \ , \quad
\nabla_i a^i=\del_1 a^1 = \frac{-6u^2-6u^4+4u^6}{U_1^2 U_2^2} \ , \quad
g_{1k}g^{kl}_{(1)} g_{l1}=\g \frac{u^2v^2}{1+u^2 v^2} \ .
\ee
\vskip-2.mm
\noindent
Upon inserting these results into eqs \eqref{F0sol} and \eqref{F1eq2} we get
\be\label{F0ntex}
F_0\big((u',v'),(u,v)\big)=1+\frac{u^3}{U_1 U_2}\D u +\Big(\frac{1+14u^2+12 u^4-6u^6}{12 U_1^2 U_2^2} +\frac{\g}{8}\frac{u^2 v^2}{1+u^2 v^2}\Big)\D u^2 +\frac{1-4u^2-6u^4}{12 U_1 U_2^3} \D v^2 \ ,
\ee
and
\be\label{F1ntex}
F_1\big((u',v'),(u,v)\big)=\frac{1+5u^2+3u^4-9u^6}{3U_1^2 U_2^2} -\frac{\g}{4} \frac{u^2 v^2}{1+u^2v^2}\ .
\ee

\section{Conclusions}

In this note, we have addressed the problem of solving the $d$-dimensional Fokker-Planck equation for completely arbitrary space and time dependent diffusion matrix and drift terms, for any initial condition, as an asymptotic expansion in the time interval $t-t_0$. As customary, we have reformulated this problem as finding the corresponding Fokker-Planck kernel $K(t,x;t_0,y)$ that corresponds to the solution for a Dirac delta-type  initial distribution. We have taken advantage of the geometrical picture which interprets the diffusion matrix  as the inverse metric on a $d$-dimensional Riemannian manifold. For time-independent diffusion matrix, the kernel $K$ then is simply related to the heat kernel of the Laplace operator on this Riemannian manifold. This heat kernel has a well-known  small-time asymptotic expansion with coefficients involving various expressions built form the Riemann curvature tensor. This expansion can be straightforwardly obtained as a perturbation series around the flat-space heat kernel. We have adapted this perturbative procedure to take into account that, at present, this Riemannian geometry depends itself on time and thus the time derivatives generate various extra terms.

We have provided the infinite set of recursive differential relations that determine the coefficient functions in this asymptotic expansion, and explicitly worked out, to full generality, the leading term and first corrections to order $t-t_0$. For the somewhat simpler case of vanishing drift force, we have also obtained the next-to-next-to-leading order corrections of order $(t-t_0)^2$, which contain many new terms as compared to the time-independent diffusion.
We have also worked out a few examples. Some of them are trivial in the sense that the time-dependence of the diffusion matrix and drift coefficients can be undone by some appropriate reparametrisation. Nevertheless, these  trivial examples provide a non-trivial consistency check of our general formulae. We also worked out a non-trivial example, mainly to illustrate that our recursive solution is well-defined and straightforward to implement. Finally, the appendix contains a pedestrian introduction to some notions of Riemannian geometry, as well as a couple of formulae about Riemann normal coordinates and the geodesic length, used in the main text.

\def\baselinestretch{1.3}
\begin{appendix}
%
\section{Appendix}

\subsection{Some elements of Riemannian geometry\label{Riemannian}}

This appendix aims at providing the reader not familiar with Riemannian geometry with a few basic formula. It is meant to be pragmatic, rather than general or precise.

A $d$-dimensional manifold ${\cal M}$ is a space on which one can define coordinates $x^i_{(r)}$, $i=1,\ldots d$ on  open sets $U_r\in {\cal M}$ (i.e.~a continuous bijection between $U_r$ and an open set in ${\bf R}^d$), such that the union of these $U_r$ covers the whole manifold ${\cal M}$ and such that on the (non-empty) overlaps $U_r \cap U_s$ there is an invertible map between the $x^i_{(r)}$ and the $x^j_{(s)}$. In particular, one can also introduce two sets of coordinates within the same (fixed) open set and one then often writes simply $x^i$ and $x^{'i}$. A trivial example is the space ${\bf R}^3$ with one open set being all of ${\bf R}^3$ with the $z$-axis removed. Then two different coordinate systems are provided by Cartesian coordinates $x,y,z$ and spherical coordinates $r,\t,\vf$. (One must exclude the $z$-axis  where $\vf$ is ill-defined.) A less trivial example is the two-dimensional sphere $S^2$ where one can use the spherical coordinates $\t,\vf$ everywhere except at the north and south-pole. To cover the full sphere one then needs to introduce other ``spherical" coordinates, say $\t',\vf'$ defined with respect to a different ``north" and ``south" pole.

On a \emph{Riemannian manifold} there is the notion of infinitesimal length $\d s$ defined as
\be\label{metricdef}
\d s^2= g_{ij}(x) \d x^i \d x^j \ ,
\ee
where the repeated indices are summed from $1$ to $d$. The coefficients $g_{ij}(x)$ form the \emph{metric tensor} at the point $x$. Note that $g_{ij}$ obviously is a non-negative, symmetric tensor. For example, on the sphere $S^2$ of radius $r_0$ one has $\d s^2= r_0^2\, \d\t^2 + r_0^2 \sin^2\t\, \d\vf^2$, so that with $x^1=\t$ and $x^2=\vf$ the non-vanishing components of the metric tensor are $g_{11}=r_0^2$ and $g_{22}=r_0^2 \sin^2\t$. The inverse metric tensor is simply denoted with upper indices $g^{ij}$ and satisfies $g_{ik}g^{kj}=\dd_i^j$. It is also symmetric and non-negative. In Riemannian geometry the metric tensor is the basic object that determines the geometric properties of the manifold. It allows us to compute the curvature properties of the manifold in every point (encoded in the curvature tensor $R^i_{\ jkl}$), define the parallel transport of vectors along any path and determine the geodesics, i.e. the paths of shortest length between two points. The components of the metric tensor depend on the coordinates used. Indeed, it is clear from \eqref{metricdef} that if one uses another set of coordinates $x^{'k}$, the infinitesimal distance will read
\be\label{changecoord}
\d s^2= g_{ij}(x) \frac{\del x^i}{\del x^{'k}} \frac{\del x^j}{\del x^{'l}} \d x^{'k} \d x^{'l} \equiv g'_{kl}(x')\d x^{'k} \d x^{'l} 
\quad\Rightarrow\quad
g'_{kl}(x')=g_{ij}(x) \frac{\del x^i}{\del x^{'k}} \frac{\del x^j}{\del x^{'l}} \ ,
\ee
and similarly for the inverse metric
\be\label{changecoordinv}
g^{'kl}(x')=g^{ij}(x) \frac{\del x^{'k}}{\del x^i} \frac{\del x^{'l}}{\del x^j} \ .
\ee
Actually, any tensor $T_{i_1\ldots i_p}^{\ \ \ \ j_1\ldots j_q}$ with $p$ lower  and $q$ upper indices transforms analogously with the required number of factors of $\frac{\del x^i}{\del x^{'k}}$ and of $\frac{\del x^{'l}}{\del x^j}$.
A scalar has no indices and takes the same value in any coordinate system: $S'(x')=S(x)$.
The infinitesimal length $\d s$ is an example of a scalar. 

It is easy to see that the partial derivatives $\del_i$ of a scalar transform as a tensor with one lower index (i.e.~a vector), but that the partial derivatives of a tensor do not transform as a tensor since second derivatives $\frac{\del^2 x^{'k}}{\del x_i\del x^j}$ are generated. These unwanted terms can be cancelled by defining a covariant derivative that in addition to $\del_i$ also involves the so-called \emph{Christoffel symbols} $\G^j_{im}$ which transform under coordinate changes in exactly the right way as to cancel the unwanted terms. They are given in terms of the first derivatives of the metric as
\be\label{Christoffel}
\G^i_{kl}=\frac{1}{2} g^{ij}\big(\del_k g_{jl}+\del_l g_{kj}-\del_j g_{kl}\big) 
\quad , \quad \G^i_{kl}=\G^i_{lk} \ .
\ee
The covariant derivative of a vector then is
\be\label{covder}
\nabla_i v^j=\del_i v^j + \G^j_{ik}v^k \ , \quad \nabla_i v_j=\del_i v_j -\G_{ij}^l v_l\ ,
\ee
while that of a tensor involves as many $\G$'s as there are indices. It follows from \eqref{Christoffel} that $\nabla_i g_{jk}=0=\nabla_i g^{jk}$~: the metric is covariantly constant.

The Riemann \emph{curvature tensor} $R^i_{\ jkl}$, the Ricci curvatur tensor ${\cal R}_{jl}$ and scalar curvatur ${\cal R}$ can be computed from these Christoffel symbols as
\be\label{curvature}
R^i_{\ jkl}=\del_k \G^i_{lj}-\del_l \G^i_{kj}+\G^i_{km}\G^m_{lj}-\G^i_{lm}\G^m_{kj} 
\quad , \quad {\cal R}_{jl}=R^i_{\ jil}
\quad , \quad {\cal R} = g^{jl}{\cal R}_{jl} \ .
\ee
$R_{ijkl}=g_{im}R^m_{\ jkl}$ has various symmetry and antisymmetry properties under exchange of its indices~:
\be\label{Rsym}
R_{ijkl}=R_{klij}=-R_{jikl} =-R_{ijlk} \ , \quad R_{ijkl}+R_{iljk}+R_{iklj}=0 \ .
\ee
In 2 dimensions it only has one independent component which can by expressed in terms of ${\cal R}$. But in 3 and higher dimensions there are more than one independent curvature components that characterise the curvature of a manifold at any given point. Note that these definitions of curvature are ``intrinsic" to the manifold and are not related to the way this manifold is possibly embedded into any higher-dimensional space. For example, a 2-dimensional torus has vanishing (intrinsic) curvature, contrary to our intuition from viewing it as embedded in ${\bf R}^3$.

The Christoffel symbols also serve to write the differential equation obeyed by a path $z^i(\l)$, parametrized by some $\l\in[0,\ell]$, that is a \emph{geodesic}, i.e. such that the distance $\int_0^\ell \d s$  between its two end-points is minimal:
\be\label{geodesiceq}
\ddot z^i +\G^i_{kl}(z)\dot z^k \dot z^l =0 \ ,
\ee
were $\dot z^i\equiv \dot z^i(\l)=\frac{\d z^i(\l)}{\d\l}$. Of course, through a given point $x$ there are infinitely many geodesics. On the 2-dimensional sphere e.g. the geodesics at the north pole are all grand circles that go from the north pole to  the south pole and back to the north pole. The \emph{geodesic length} $\ell(x,y)$ is the length of the geodesic that goes through $x$ and $y$. Generically there is only one such geodesic from $x$ to $y$, but in special cases it may happen that there are several of equal length. Of course, if $x$ and $y=x+\d x$ are infinitesimally close, the geodesic is just the straight line $\d x^i$ and $\ell^2(x,x+\d x)=\d s^2=g_{ij}(x)\d x^i \d x^j$.

\subsection{Riemann normal coordinates\label{app2}}

One may use the geodesics through a given point $y$ to define new coordinates, called \emph{Riemann normal coordinates}. If one takes $d$ geodesics $z^i_{(k)}(\l),\ k=1,\ldots d$, through $y$ with linearly independent tangent vectors $\dot z^i_{(k)}(0)$ at $y$ and defines the new coordinates $\xi^k$ as increasing along this $k^{\rm th}$ geodesic and being equal to its length as measured from $y$ then, obviously, in these new coordinates the geodesics are simply affine functions, i.e. $\ddot \xi^k=0$. Thus, in these coordinates the $\G^i_{mn}$ vanish in $y$ (as do all symmetrized derivatives $\del_{(j_1} \ldots \del_{j_k} \G^i_{mn)}$). This implies that the first derivatives of the metric at $y$ vanish in these coordinates, while the second derivatives can be directly expressed in terms of the curvature tensor.\footnote{
In general relativity these are called locally inertial coordinates and correspond locally to the coordinate system of a freely falling observer.
} 
Upon choosing the tangent vectors $\dot z^i_{(k)}(0)$ appropriately, the metric close to $y$ then has the following form (see e.g.~the appendix of \cite{BF})
\be\label{metricnormalexp}
g_{ij}(\xi)= \dd_{ij}-{1\over 3} R_{ikjl}\xi^k \xi^l -{1\over 6}R_{ikjl;m}\xi^k \xi^l \xi^m 
+\ \Big[{2\over 45} R_{ikrl}R^r_{\ mjn} 
-{1\over 20}R_{ikjl;mn}\Big] \xi^k \xi^l \xi^m \xi^n +{\cal O}(\xi^5) \, ,
\ee
where $(\ldots)_{;m}$ denotes a covariant derivative, and all curvature tensors are to be evaluated at $\xi=0$ (i.e.~at $y$). The inverse metric is easily seen to be
\be\label{inversemetricexp}
g^{ij}(\xi)= \dd_{ij}+{1\over 3} R_{ikjl}\xi^k \xi^l +{1\over 6}R_{ikjl;m}\xi^k \xi^l \xi^m 
+\ \Big[{1\over 15} R_{ikrl}R^r_{\ mjn} 
+{1\over 20}R_{ikjl;mn}\Big] \xi^k \xi^l \xi^m \xi^n +{\cal O}(\xi^5) \ .
\ee
Note that in these coordinate one has
\be\label{gx}
g_{ij} \xi^j=\xi^i \quad , \quad g^{ij} \xi^j = \xi^i \ ,
\ee
since all other terms involve symmetric products of the coordinates $\xi$ contracted with antisymmetric curvature tensors.  From the definition of these coordinates it follow that the geodesic length between the point $y$ which is $\xi=0$ and the point $x$ corresponding to $\xi$ is
\be\label{geonormal}
\ell^2(0,\xi)= \sum_i \xi^i \xi^i \ .
\ee
In particular, 
\be\label{georelapp}
\frac{\del}{\del \xi^i} \ell^2(0,\xi) = 2 \xi^i 
\quad\Rightarrow\quad
g^{ij}(\xi) \frac{\del \ell^2(0,\xi)}{\del \xi^i}  \frac{\del \ell^2(0,\xi) }{\del \xi^j} = 4 \ell^2(0,\xi) \ .
\ee

Some further useful relations valid at the origin of the normal coordinates, i.e. at $\xi=0$ are the following~:
\be\label{Christoffelandder}
{\rm at}\ \xi=0 \ : \quad \G^i_{kl}=0 \ , \quad \del_k\G^i_{lj}+\del_l\G^i_{kj}+\del_j\G^i_{kl}=0 \quad \Rightarrow\quad
\del_k\G^i_{jl}=\frac{1}{3}(R^i_{\ jkl}+R^i_{\ lkj}) \ .
\ee
Denoting covariant derivatives of any tensor as $\nabla_k T_{ij}\equiv T_{ij;k}$ and  $\nabla_k\nabla_l T_{ij}\equiv T_{ij;kl}$ (with the indices in this order) etc, one has $\del_k T_{ij}=T_{ij;k}$ (at $\xi=0$), as well as
\ba\label{delnablaT}
{\rm at}\ \xi=0 \ : \qquad  g^{kl}\del_k \del_l T_{ij}&=&g^{kl}T_{ij;lk} - \frac{1}{3} \cR^m_{\ i} T_{mj}  - \frac{1}{3} \cR^m_{\ j} T_{im} \ ,
\nonumber\\
\del_k\del_l T^i_{\ i}&=&T^i_{\ i;lk} \ ,
\nonumber\\
\del_k\del_i T^i_{\  j} &=&T^i_{ \ j;ik} +\frac{1}{3} \cR^i_{\ k} T_{ij} -\frac{1}{3} ( R_{rskj}+R_{sjrk})T^{rs} 
\nonumber\\
&=& T^i_{ \ j;ki}  -  \frac{2}{3} \cR^i_{\ k} T_{ij}  + \frac{1}{3} (R_{sjrk}+R_{skrj})T^{rs} \ .
\ea

\subsection{Some formulae for the geodesic length and its derivatives\label{app3}}

The geodesic from $y$ to $x$ is obtained by solving the geodesic equation \eqref{geodesiceq} and adjusting the initial condition $\dot z^i(0)$ such that  $z^i(s)=x^i$ for $s=\ell(x,y)$. Expanding $z^i(s)$ in a Taylor series in $s$, $z^i(s)=y^i+s \dot z^i(0) +\frac{s^2}{2} \ddot z^i(0) +\frac{s^3}{6} \dddot z^i(0) +\ldots$ and using the geodesic equation and its derivatives to express $\ddot z^i(0)$ and all higher derivatives through products of $\dot z^k(0)$ one gets
\be\label{zieq}
z^i(s)=y^i+ s \dot z^i(0) -\frac{s^2}{2} \G^i_{jk} \dot z^j(0)\dot z^k(0) - \frac{s^3}{6} \big( \del_l\G^i_{jk} -2 \G^i_{mk}\G^m_{jl}\big) \dot z^l(0)\dot z^j(0)\dot z^k(0) + \ldots \ .
\ee
Here all $\G^i_{jk}$ and $\del \G^i_{jk}$ are taken at $s=0$, i.e. at $y$. If we let $\eta^i(s)=s \dot z^i(s)$ and $s=\ell(x,y)$ this can be rewritten as
\be\label{epseta}
\e^i\equiv x^i-y^i=\eta^i-\frac{1}{2} \G^i_{jk}\eta^j\eta^k- \frac{1}{6} \big( \del_l\G^i_{jk} -2 \G^i_{mk}\G^m_{jl}\big)\eta^l\eta^j\eta^k + \ldots \ .
\ee
As it stands, this relation is for one fixed pair $(x,y)$. But one can repeat this, keeping $y$ fixed and varying $x$, so that this provides new coordinates $\eta^i$ around the point $y$, obtained by inverting the previous formula :
\be\label{etaeps}
\eta^i = \e^i+\frac{1}{2} \G^i_{jk}\e^j\e^k + \frac{1}{6} \big( \del_l\G^i_{jk} + \G^i_{mk}\G^m_{jl}\big)\e^l\e^j\e^k + \ldots \ .
\ee
The construction of these coordinates $\eta^i$ ressembles the one of the normal coordinates outlined above. Indeed, the $\eta^i$ are the normal coordinates around $y$ provided $g_{ij}(0)=\dd_{ij}$. To see this, note
that it follows from our construction of the geodesics $z^i(s)$ that $g_{ij}(y) \dot z^i(0)\dot z^j(0)= \frac{\d s^2}{\d s^2}=1$, which implies 
\be\label{getaeta}
g_{ij}(y) \eta^i\eta^j=\ell^2(x,y) \ .
\ee
The ``true" normal coordinates $\xi$ around $y$ can then be obtained by ``diagonalising" the fixed positive symmetric matrix $g_{ij}(y)$ as  
\be\label{truenormal}
g_{ij}(y) = \sum_k O_i^{\ k}  \, O_j^{\ k}\quad , \quad \xi^k= \eta^i \, O_i^{\ k} \ .
\ee
However, in the remainder of this appendix, we continue to work directly with the $\eta^i$.

Inserting \eqref{etaeps} into \eqref{getaeta} yields a formula for the geodesic distance between $x$ and $y$ written as an expansion in the coordinate difference $\e^i=x^i-y^i$ :
\be\label{geodistcoord}
\ell^2(x,y)=g_{ij}\e^i\e^j +\frac{1}{2} \del_k g_{ij} \e^k\e^i\e^j 
+  \big( \frac{1}{6} \del_i\del_j g_{lk} -\frac{1}{12} g_{nm}\G^n_{ij}\G^m_{kl} \big)\e^i\e^j\e^k\e^l +\ldots \ ,
\ee
where, again, all metrics, Christoffel symbols and their derivatives are evaluated at $y$.
It follows that
\be\label{deliell}
\frac{\del\ell^2(x,y)}{\del x^i}=2g_{ij}\e^j +\frac{3}{2} \del_{(k} g_{ij)} \e^j\e^k 
+  \big( \frac{2}{3} \del_{(i}\del_j g_{lk)} -\frac{1}{3} g_{nm}\G^n_{(ij}\G^m_{kl)} \big)\e^j\e^k\e^l +\ldots \ ,
\ee
(where $a_{(i_1\ldots i_n)}$ denotes symmetrization in the indices, e.g.~ $a_{(ij)}=\frac{1}{2}(a_{ij}+a_{ji})$), and
\be\label{deldeliell}
\frac{\del^2\ell^2(x,y)}{\del x^i \del x^j}=2g_{ij} +3 \del_{(k} g_{ij)} \e^k 
+  \big( 2 \del_{(i}\del_j g_{lk)} -g_{nm}\G^n_{(ij}\G^m_{kl)} \big)\e^k\e^l \ .
\ee
Using also the expansion of the metric $g_{ij}(x)$ and the inverse metric $g^{ij}(x)$ around $y$
\ba\label{metricexp}
g_{ij}(x)&=&g_{ij}+\del_p g_{ij}\e^p+\frac{1}{2}\del_p\del_q g_{ij}\e^p\e^q+\ldots
\\
g^{ij}(x)&=&g^{ij}-g^{il}\del_p g_{lk}g^{kj}\e^p-\frac{1}{2}g^{il}\del_p\del_q g_{lk}g^{kj}\e^p\e^q +g^{il}\del_p g_{lm}g^{mk}\del_q g_{kn}g^{nj}\e^p\e^q+\ldots
\ea
one  gets
\be\label{gijdeliell}
g^{ij}(x)\frac{\del\ell^2(x,y)}{\del x^j}
=2\e^i-\G^i_{kl}\e^k\e^l -\frac{2}{3} \del_k \G^i_{pq}\e^k\e^p\e^q 
+\frac{1}{3}\G^i_{kl}\G^l_{pq} \e^k\e^p\e^q
+\ldots \ ,
\ee
and, multiplying with $\frac{\del\ell^2}{\del x^i}$ from \eqref{deliell}, one can 
check, up to and including terms of order 4 in $\e$, that one indeed has 
\be\label{gellellell}
g^{ij}(x) \frac{\del\ell^2(x,y)}{\del x^i} \frac{\del\ell^2(x,y)}{\del x^j} = 4 \ell^2(x,y) \ ,
\ee
in agreement with \eqref{georelapp}.
Finally, we find
\ba\label{gellsecond}
g^{ij}(x)\frac{\del^2\ell^2(x,y)}{\del x^i \del x^j}
&=& 2d + 2 g^{rs}\G^m_{rs} g_{mk}\e^k
+\frac{1}{3}g^{ij}(\del_i\del_j g_{lk}-2\del_k\del_l g_{ij}+4\del_i\del_k g_{jl})\e^k\e^l
\nonumber\\&&
-g^{ij}g_{mn}\G^n_{(ij}\G^m_{kl)} \e^k\e^l 
+2\del_l g^{rs}g_{kj}\G^j_{rs}\e^k\e^l +\ldots \ ,
\ea
\vskip-3.mm
\noindent
and
\vskip-3.mm
\noindent
\ba\label{gammaterm}
g^{rs}(x)\G^j_{rs}(x)\del_j\ell^2(x,y)&=&2 g^{rs}\G^m_{rs} g_{mk}\e^k
+g^{ij}(2\del_i\del_k g_{jl}-\del_k\del_l g_{ij})\e^k\e^l
\nonumber\\
&&-g^{ij}g_{mn}\G^n_{ij}\G^m_{kl} \e^k\e^l 
+2\del_l g^{rs}g_{kj}\G^j_{rs}\e^k\e^l +\ldots \ .
\ea
(Recall that all quantities on the right-hand side are evaluated at $y$.)
Combining the last two equations and rearranging a bit we get
\ba\label{Laplell}
\D_x \ell^2(x,y)&=&g^{ij}(x)\del_i\del_j\ell^2(x,y)-g^{rs}(x)\G^j_{rs}(x)\del_j\ell^2(x,y)
\nonumber\\
&=&2d -\frac{2}{3}g^{ij}\Big[\frac{1}{2}\big(\del_i\del_l g_{jk}+\del_j\del_k g_{il}
-\del_i\del_j g_{kl}-\del_k\del_l g_{ij}\big) +g_{nm}\G^n_{il}\G^m_{jk}-g_{nm}\G^n_{ij}\G^m_{kl}\Big]\, \e^k\e^l +\ldots
\nonumber\\
&=&2d -\frac{2}{3} {\cal R}_{kl}\e^k\e^l +\ldots \ .
\ea
It is not a surprise that the final result can be simply expressed in terms of the Ricci curvature tensor. Indeed, $\D \ell^2$ is a scalar quantity and can be computed in any coordinate system. The expression in normal coordinates $\xi$ around $y$ can be found e.g. in \cite{BF}, up to and including terms of order 4 in $\xi$, and can be entirely expressed in terms of the curvature tensor and its covariant derivatives. Up to order 2 in $\e$ it coincides with the 
$2d -\frac{2}{3} {\cal R}_{kl}\e^k\e^l $ found in \eqref{Laplell}.

\end{appendix}

\noindent

\noindent

\vskip-8.mm
\newpage


\end{document}